\documentclass[
reprint,
superscriptaddress,
amsmath,
amssymb,
aps,
prx,
twocolumn]{revtex4-2}
\usepackage[utf8]{inputenc}
\usepackage{natbib}
\usepackage{physics}
\usepackage{braket}
\usepackage{amsmath}
\usepackage{bbold}
\usepackage{verbatim}
\usepackage{amssymb}
\usepackage{amsthm}
\usepackage{graphicx} 
\usepackage{caption}
\usepackage{subcaption}
\usepackage{booktabs}
\usepackage{enumitem}
\usepackage[dvipsnames]{xcolor}
\usepackage{amsmath}
\usepackage[english]{babel}
\usepackage{soul}
\usepackage{csquotes}
\usepackage{enumitem}
\usepackage{amsfonts}
\usepackage{tcolorbox}
\usepackage{enumitem}
\usepackage{adjustbox}
\usepackage{standalone}
\usepackage{tikz}
\usepackage{bbm}
\usepackage{breqn}
\usepackage{float}
\usepackage{algorithm}
\usepackage{algpseudocode}
\usepackage{flafter}
\usepackage{wrapfig}
\usepackage{ragged2e}
\usepackage[justification=justified, font=small]{caption}

\usetikzlibrary{quantikz}

\newtheorem{theorem}{Theorem}
\newtheorem{lemma}{Lemma}

\usepackage[section]{placeins}
\usepackage{afterpage}
\definecolor{mpq}{RGB}{40, 73, 93}
\usepackage[colorlinks=true, linkcolor=purple, citecolor=teal, urlcolor=mpq]{hyperref}

\newcommand{\Id}{\mathbbm{1}}
\newcommand{\map}[1]{\mathcal{#1}}

\newcommand{\fab}{f_{\alpha_i\beta_i}(r_i,s_i)}
\newcommand{\gab}{g_{\alpha_i\beta_i}(r_i,s_i)}
\newcommand{\ggd}{g_{\gamma_i\delta_i}(r_i,s_i)}
\newcommand{\vf}{\textit{\bf{f}}}
\newcommand{\vg}{\textit{\bf{g}}}

\newcommand{\balpha}{\vec{\alpha}}
\newcommand{\bbeta}{\vec{\beta}}
\newcommand{\bgamma}{\vec{\gamma}}
\newcommand{\bdelta}{\vec{\delta}}
\newcommand{\bchi}{\boldsymbol{\chi}}
\newcommand{\chiab}{\chi_{\balpha\bbeta}}
\renewcommand{\vr}{\vec{r}}
\newcommand{\vs}{\vec{s}}
\newcommand{\br}{\vec{r}}
\newcommand{\bs}{\vec{s}}
\newcommand{\bj}{\textbf{j}}

\newcommand{\bG}{\boldsymbol{G}}

\newcommand{\mapd}[1]{\mathcal{#1}^{\dagger}}

\newcommand{\scalingeq}{e^{{\mathcal{O}}(d \log(1/\epsilon))} \cdot \map{O}\left(\log\left(\frac{n}{\delta}\right)\right)}
\newcommand{\scalingdiamond}{n^{\mathcal{O}(d)}e^{{\mathcal{O}}( \log(1/\epsilon))} \cdot \map{O}\left(\log\left(\frac{1}{\delta}\right)\right)}
\newcommand{\scalingdiamondpp}{n^{\mathcal{O}(d)}e^{{\mathcal{O}}( \log(1/\epsilon))+q} \cdot \map{O}\left(\log\left(\frac{1}{\delta}\right)\right)}

\newcommand{\scalingeqpp}{e^{{\mathcal{O}}(d \log(1/\epsilon))} \cdot \map{O}\left(n\log\left(\frac{n}{\delta}\right)\right)}
\newcommand{\scalingeqppexp}{e^{{\mathcal{O}}(d \log(1/\epsilon))} \cdot \map{O}\left(ne^q\log\left(\frac{n}{\delta}\right)\right)}

\renewcommand{\map}[1]{\mathcal{#1}}

\begin{document}
		
    \title{Efficient Characterization of Coherent and Correlated \\ 
    Low-Degree Noise in Layers of Gates}

    \newcommand{\mpq}{Max Planck Institute of Quantum Optics, 85748 Garching, Germany}
    \newcommand{\mcqst}{Munich Center for Quantum Science and Technology (MCQST), 80799 \ Munich, Germany}
    \newcommand{\padua}{Dipartimento di Fisica e Astronomia \enquote{G. Galilei} \& Padua Quantum Technologies Research Center, Università degli Studi di Padova, 35131 Padua, Italy}

\author{Marianna Crupi}
\email{marianna.crupi@mpq.mpg.de}
\affiliation{\mpq}
\affiliation{\mcqst}
\author{J. Ignacio Cirac}
\affiliation{\mpq}
\affiliation{\mcqst}
\author{Flavio Baccari}
\affiliation{\padua}

\begin{abstract}
We present a quantum process-tomography protocol based on a low-degree ansatz for the quantum channel, i.e. when it can be expressed as a fixed-degree polynomial in terms of Pauli operators. We demonstrate how to perform tomography of such channels with a logarithmic amount of effort relative to the size of the system, by employing random state preparation and measurements in the Pauli basis. We extend the applicability of the protocol to channels consisting of a layer of quantum gates with a polylogarithmic number of non-Clifford gates, followed by a low-degree noise channel. Rather than inverting the layer of quantum gates on the hardware—which would introduce additional errors—we instead carry out the inversion in classical postprocessing, while adding to the sample complexity a factor at most polynomial in system size. Numerical simulations support our theoretical findings and demonstrate the feasibility of our method.
\end{abstract}
	
\maketitle

\section{Introduction}
Accurate noise characterization is a critical prerequisite for the reliable use of current quantum devices and the development of future quantum technologies. Understanding the main sources of imperfections in quantum operations is essential for designing effective error mitigation and correction techniques~\cite{kim2023evidence,endo2021hybrid,seif2023shadow,terhal2015quantum}, as well as for assessing the capabilities of quantum hardware~\cite{blume-kohout2017demonstration,RB}. 

An effective noise characterization scheme should satisfy several key criteria. First, it must be scalable, in terms of both experimental implementation and the computational resources required to process the collected data. Second, the noise model should be sufficiently comprehensive to capture the complexity of realistic noise processes and ensure broad applicability across various platforms. In particular, it should account for both coherent errors and crosstalk, even with arbitrary device connectivity. \par
A wide range of noise characterization methods have already been developed, each navigating trade-offs between scalability and generality. Standard approaches like quantum process-tomography and gate-set tomography aim to fully reconstruct quantum channels but quickly become impractical due to their exponential resource demands~\cite{poyatos1997complete, PRXQuantum.2.040338,nielsen2021gate,merkel2013self, chuang1997prescription, ProjectedLSQPT}. More scalable alternatives, such as randomized benchmarking and shadow-tomography, enable efficient estimation of specific noise properties but sacrifice completeness in the reconstructed information~\cite{gambetta2012characterization, levy2024classical, mckay2020correlated, helsen2022general, huang2023learning, non_linear, RB, ShadowEstimation, st,sarovar2020detecting}. A promising intermediate strategy seeks to balance scalability and completeness by imposing assumptions on the noise model. Examples include Pauli-diagonal noise models, perturbative ans\"atze, or low-degree models~\cite{kaufmann2025characterization,harper2020efficient,rouze2023efficient,population_recovery,govia2020bootstrapping,torlai2023quantum,wadhwa2024agnostic,learning_lowdegree}. While each approach suits specific use cases, they come with limitations: some require fine-tuned noise tailoring (e.g., Pauli twirling), others lack rigorous recovery guarantees, and some are experimentally impractical due to high circuit complexity. \par
In this work, we introduce a tomography method that maintains both scalability and completeness while overcoming the limitations of previous approaches. Our focus is on the low-degree quantum channel ansatz~\cite{learning_lowdegree}, where error events act nontrivially on at most $d$ qubits simultaneously. This ansatz captures important physical features such as coherent errors and crosstalk, and it generalizes to devices with arbitrary connectivity. We address the scalability challenges of earlier tomography schemes~\cite{poyatos1997complete, PRXQuantum.2.040338,nielsen2021gate,merkel2013self, chuang1997prescription, ProjectedLSQPT} by developing a reconstruction algorithm with logarithmic sample complexity, relying only on single-qubit state preparations and measurements (SPAMs). As a result, our method is both theoretically efficient and experimentally practical. 
\begin{figure*} 
  \centering
  \includegraphics[width=\textwidth]{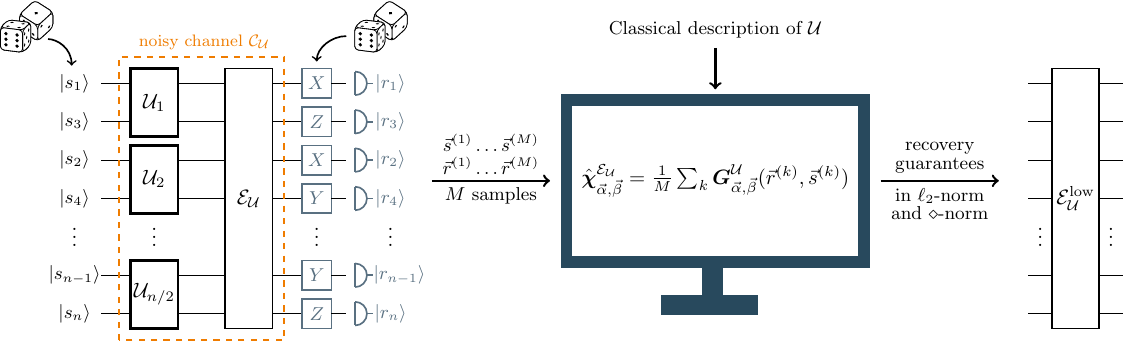}
  \caption{Schematic summary of the proposed tomography protocol. A random set of single-qubit Pauli eigenstates is initialized, evolved through the noisy channel $\map{C}_{\map{U}}$ and measured in a random Pauli basis. The classical outcome of the measurements is then sent to a classical computer that uses them alongside the classical description of $\map{U}$ (e.g., its matrix representation in the computational basis) to compute the estimated process matrix $\bchi^{\map{E}_{\map{U}}}$ by averaging over the function $\boldsymbol{G}^{\map{U}}$ which converges entrywise to $\bchi^{\map{E}_{\map{U}}}$. The final output of the protocol is then the low-degree approximation of the noisy part of the implementation $\map{E}_{\map{U}}$.}
  \label{fig:protocol}
\end{figure*}

The low-degree ansatz is particularly well suited for describing weak noise channels. It can also model the noisy implementation of one or a few two-qubit gates, including potential crosstalk with idle qubits. However, when an entire layer of such gates is applied, the resulting channel ceases to be low degree, regardless of the noise strength. To address this, we model the noisy implementation as an ideal gate layer followed by a low-degree noise channel. Recovering the noise component alone requires inverting the ideal gate layer—a process that can introduce additional noise if performed on hardware. Instead, we demonstrate that in certain relevant cases, our protocol can eliminate the effect of ideal gates through classical postprocessing, avoiding extra experimental overhead. This extension broadens the applicability of our protocol and makes it a versatile tool for characterizing noise in near-term quantum devices, while only adding an overhead at most polynomial in system size. \par
This paper is organized as follows. In Sec.~\ref{problem_setting}, we introduce our noise characterization framework. Section~\ref{rm} presents the tomography protocol for low-degree channels and provides recovery guarantees.
In Sec.~\ref{postprocessing} we show how to adapt the protocol to eliminate gate layers via postprocessing. Section~\ref{numerics} presents numerical results under specific noise models. Finally, Sec.~\ref{conclusion} summarizes our findings and discusses future directions.

\section{Problem setting}\label{problem_setting}
We consider the application of a noisy unitary on $n$ qubits, and the goal is to completely recover the noise component. For this we assume that certain laboratory operations—specifically, state preparation and measurement—can be performed without error. In particular, we assume the ability to prepare product states of Pauli eigenstates $(\sigma_x,\sigma_y,\sigma_z)$ and to perform measurements in these bases. While this assumption may not hold in all experimental settings, it is well justified in several quantum computing platforms—for example, when high-fidelity measurements in the computational basis and accurate single-qubit Clifford gates are available. \par
We now recall some standard concepts from quantum process-tomography, in particular those relevant for gate-error characterization, which will be useful for the development of our protocol.\par
Let us represent the noisy implementation of a unitary by the corresponding quantum channel,
\begin{equation}
    \map{U} \mapsto \map{C}_{\map{U}}.
\end{equation}
In the general case this channel can be many-body and gate-dependent, including correlated and coherent noise models. \\
More specifically, we would like to characterize the error component of the noisy implementation $\map{C}_{\map{U}}$,
\begin{equation}\label{E}
    \map{E}_{\map{U}} := \map{C}_{\map{U}} \circ  \map{U}^\dagger,
\end{equation}
where we use the subscript $\map{U}$ to indicate the possible gate dependence of the noise \footnote{Notice that there is a degree of arbitrariness in defining $\map{E}_{\map{U}} := \map{C}_{\map{U}} \circ  \map{U}^\dagger$ or $\map{E}'_{\map{U}} := \map{U}^\dagger \circ \map{C}_{\map{U}}$. The two representation are connected via $\map{E}_{\map{U}} = \map{U} \circ \map{E}'_{\map{U}} \circ \map{U}^\dagger$}. However, we will omit the subscript where there is no chance of misinterpretation. 
\par 
Characterizing a generic quantum channel requires resources that scale exponentially with system size~\cite{chuang1997prescription}. To enable an efficient representation of the noise in our implementation, we must therefore make further assumptions. We adopt a physically motivated one: the error channel is a perturbative expansion around the identity channel, corresponding to the case of no noise. More specifically, we define the Kraus operators of the error channel as in Eq.~\eqref{E},
\begin{equation}
    \label{kraus_rep}
    \map{E}(\rho) = \sum_{k} E_k \rho E_k^{\dagger},
\end{equation}
and we assume that they act nontrivially on at most a fixed number $d$ of qubits at a time. That is, the Pauli decomposition of Kraus operators,
\begin{equation}\label{eq:Krausexp}
E_k = \sum_{\balpha \in I_d} e_{k\balpha} \sigma_{\balpha}    
\end{equation}
only takes nonzero values in the set $I_d :=  \{\balpha \in \lbrace 0,x,y,z \rbrace^{n}| \sigma_{\balpha}$ acts nontrivially on at most $d$ qubits$\}$, where with $\sigma_0$ we mean the identity operator. This set has cardinality $D = \sum_{i=0}^d 3^d\binom{n}{d}$ and is therefore polynomial in $n$.
A convenient way to parametrize channels of this kind is through the \textit{process matrix} representation,
\begin{equation}\label{channel_assump}
    \map{E}(\rho) = \sum_{\balpha,\bbeta \in I_d} \chi_{\balpha\bbeta}\sigma_{\balpha}\rho \sigma_{\bbeta} \, 
\end{equation}
where we have made use of the expansion in~\eqref{eq:Krausexp} and defined $\chi_{\balpha\bbeta} = \sum_k e_{\bbeta k}^*e_{k\balpha}$ . While being exponentially large in the general case, channels following our ansatz are represented by process matrices of polynomial size: $ \chi_{\balpha\bbeta} = (D \times D)$.
They capture the intuition of stochastic errors occurring at once only on few qubits and are often referred to as $d$-degree (or \textit{low-degree}) Pauli channels, a class that has already been explored in the literature~\cite{learning_lowdegree}. To keep the model broadly applicable, we make no assumptions about qubit connectivity, allowing it to apply to quantum devices of arbitrary layout. 
In Appendix~\ref{phys} we show that this ansatz well represents the error channels in two physically motivated settings. First, we consider weak single-qubit errors and show that the noise channel can be well approximated by a low-degree expansion provided the error probability scales at most as $\mathcal{O}(\sqrt{d/n-\log(1/\epsilon)/n})$, where $\epsilon$ denotes the target precision of the reconstruction protocol. Second, we analyze weak spurious two-qubit interactions and find that the ansatz remains accurate when the interaction strength is at most $\mathcal{O}(\epsilon^{1/d}/n^{2})$. These examples support the idea that low-degree channels capture realistic error models in regimes of practical interest (e.g. weak noise). \\
Importantly we notice that the error component $\map{E}$ is a low-degree quantum channel, while the whole noisy implementation $\map{C}$ might not be. For instance, if we consider a layer of Pauli gates, it will be represented by full weight Pauli strings and so will its noisy implementation. This observation motivates our work. With respect to previous techniques~\cite{learning_lowdegree} we provide contributions that address two challenges:
\begin{itemize}
\color{black}
    \item first we define an efficient and experimentally friendly tomography scheme for low-degree quantum channels;
    \item then we adapt the protocol to reconstruct the channel $\map{E}$ while only having experimental access to $\map{C}$.
\end{itemize}  
We seek a tomography protocol that scales at most polynomially with system size and relies only on single-qubit state preparation and measurement. This requirement is consistent with our assumption that single-qubit operations are approximately noiseless, ensuring that the state preparation and measurement procedures do not introduce additional noise into the system. \\
The proposed tomography method is summarized in Fig.~\ref{fig:protocol}. By employing randomized Pauli state preparation and measurement, we show that the error channel can be learned to a desired accuracy in $\ell_2$-distance  using only a logarithmic number of samples. In addition, we relate the diamond distance between two channels to the $\ell_1$-distance between their process matrices, and show that the resulting reconstruction cost in diamond distance remains polynomial in the system size. Furthermore, we demonstrate that the noisy component of the implementation (Eq.~\eqref{E}) in many relevant cases can be extracted entirely in classical postprocessing—avoiding the need to invert the unitary layer on hardware, which would itself be an additional source of error. This inversion comes with an overhead at most polynomial in sample complexity. In what follows, we first introduce the tomography protocol and then adapt it to include the classical postprocessing part.

\section{Tomography protocol} \label{rm}
We consider a randomized Pauli measurement tomography scheme, which has been widely successful in the context of state-tomography~\cite{shadow}. For each qubit, we sample randomly an initial Pauli eigenstate and a measurement basis from the Pauli basis. We denote the input and output states by $\ket{\vs}$ and $\ket{\vr}$, taking the six possible values $\{\ket{+},\ket{-},\ket{i},\ket{-i},\ket{0},\ket{1}\}$. The probability $p(\vr,\vs)$ of measuring an output $\ket{\vr}$, given an input $\ket{\vs}$ can be written as
\begin{align} \label{F}
     p(\vr,\vs) & = \frac{1}{18^n} \bra{\vr} \map{E}(\ketbra{\vs}{\vs})\ket{\vr} \nonumber \\
     & = \sum_{\balpha\bbeta} \chi_{\balpha\bbeta}F_{\balpha\bbeta}(\vr,\vs) ,
\end{align}
by introducing the function 
\begin{equation}
F_{\balpha\bbeta}(\vr,\vs) = \frac{1}{18^n} \bra{\vr}  \sigma_{\balpha} \ketbra{\vs}{\vs} \sigma_{\bbeta}\ket{\vr} ,
\end{equation}
with the $1/18^n$ prefactor chosen to ensure the overall normalization $\sum_{\vr,\vs}p(\vr,\vs) = 1$. 
We can interpret $\boldsymbol{F}$ as a matrix whose rows are indexed by the process matrix entries $\balpha,\bbeta$ and whose columns are indexed by the possible input and output states $\vr,\vs$. Following this formalism, we can regard both $\chi_{\balpha\bbeta}$ and $p(\vec{r},\vec{s})$ as vectors indexed by $\balpha,\bbeta$ and $\vec{r},\vec{s}$, respectively. In this notation, Eq.~\eqref{F} can be rewritten as a matrix–vector product,
\begin{equation}
\vec{p} = \boldsymbol{F}\vec{\chi}.
\end{equation}
To recover $\vec{\chi}$ from $\vec{p}$, we need to define a left inverse of $\boldsymbol{F}$, which we denote by $\boldsymbol{G}$. Any matrix $\boldsymbol{G}$ satisfying this condition allows us to write
\begin{equation}
\vec{\chi} = \boldsymbol{G}\vec{p},
\end{equation}
or, more explicitly,
\begin{equation} \label{G}
    \chi_{\balpha\bbeta} = \sum_{\vec{r},\vec{s}} p(\vec{r},\vec{s})\, G_{\balpha\bbeta}(\vec{r},\vec{s}) = \mathbb{E}[G_{\balpha\bbeta}] \, .
\end{equation}
Hence, computing the value $G_{\balpha\bbeta}(\vec{r},\vec{s})$ for each snapshot allows one to recover the corresponding entry $\chi_{\balpha\bbeta}$ by averaging over sufficiently many samples. An example of such a function is provided in the well-known shadow-tomography protocol~\cite{shadow,st}, where a specific choice of inverse is made. In particular, the inverse is taken to be
$\boldsymbol{G} := (\boldsymbol{F}^{\dagger}\boldsymbol{F})^{-1}\boldsymbol{F}^{\dagger}$.
We will later provide more details on this choice in the context of the shadow-tomography protocol. \par
The convergence to the actual value of $\chiab$ will scale with the variance of $G_{\balpha\bbeta}$ divided by the total number of measurements. Therefore, the key ingredient to obtain an efficient tomography protocol is to control the scaling of the variance of $G_{\balpha\bbeta}$ with the system size~\cite{shadow,st}. In the following we show how to derive a choice of $\boldsymbol{G}$ whose variance remains constant with $n$ for every entry of $\bchi$, if evaluated on snapshots coming from low-degree channels of the form~\eqref{channel_assump}.  \par
We make use of the fact that, for product input and output states, the function $\boldsymbol{F}$ takes a product form
\begin{equation} \label{F_gen}
F_{\balpha\bbeta}(\vr,\vs) = \prod_{i=1}^n \fab    \, ,
\end{equation}
where
\begin{equation} \label{f_ab}
    \fab = \frac{1}{18} \bra{r_i}\sigma_{\alpha_i} \ketbra{s_i}{s_i} \sigma_{\beta_i}\ket{r_i}    ,
\end{equation}
and index $i$ refers to the $i$-th qubit. We restrict to functions taking the same product structure $\bG_{\balpha \bbeta} = \prod_i g_{\alpha_i \beta_i}$, where $g_{\alpha_i \beta_i} = (f_{\alpha_i \beta_i})^{-1}$. More specifically we ask for
\begin{equation} \label{inv}
    \sum_{r_i,s_i} \ggd \fab = \delta_{\alpha_i\gamma_i}\delta_{\beta_i\delta_i}.
\end{equation}
A well-known example of such a construction, as mentioned previously, is provided by the shadow-tomography protocol~\cite{shadow,st}. In that case the single-qubit function $\gab$ is given by
\begin{equation}\label{g_shh}
     g^{\mathrm{sh}}_{\gamma_i\delta_i}(r_i,s_i) = \Tr[\ket{\phi_{\gamma_i}}\bra{\phi_{\delta_i}} D(r_i,s_i)],
\end{equation}
with $\ket{\phi_{\gamma_i}} = (\sigma_{\gamma_i}\otimes\Id) \ket{\phi^{+}}$, $\ket{\phi^{+}} = (\ket{00}+\ket{11})/\sqrt{2}$ and $D(r_i,s_i) = \map{D}_{1/3}^{-1}(\ketbra{r_i})\otimes\map{D}_{1/3}^{-1}(\ketbra{s_i})$, where $\map{D}_{1/3}^{-1}(A) = 3A - \Id$. As already noted in previous work, the application of this function to process-tomography often leads to variance that grows exponentially with system size~\cite{st}. In Appendix~\ref{others} we show that this also applies to the estimation of the entries of the process matrix. In contrast, our construction in the next Section provides a bounded-variance alternative tailored to low-degree channels. \par
Building on this observation, we note that---as already discussed in Refs.~\cite{dual,optimizing_caprotti,dualframe_Fischer}--the choice of the inverse map $D(r_i,s_i)$ is not unique, and different choices can lead to exponentially different variances~\cite{dual}. We will exploit this freedom to identify a form of $\boldsymbol{G}$ that minimizes the variance specifically for low-weight error channels. \\
We use our formalism to cast the problem into a quadratic minimization. Let us think of the single-qubit function $\vf := \fab$ as a rectangular matrix of dimension $(36\times16)$ with rows indexed by $r_i,s_i$ and columns indexed by $\alpha_i, \beta_i$. Given the dimension count there will be at least $20$ vectors in the left kernel of $\vf$. We denote $\mathbf{K}$ the matrix such that $K_j(r_i,s_i)$ is the $j$th element of such vectors. For every choice of $\vg$ then we have that $\vg + \vec{x}^{\mathrm{T}}\mathbf{K}$ is also a valid choice of inverse, with $\vec{x}$ being a vector of $20$ free parameters.\par
We can now choose $\vec{x}$ to get the minimum variance of the function $\bG$, expressed as
\begin{align} 
    \text{Var}&[G_{\bgamma\bdelta}] = \mathbbm{E}[|G_{\bgamma\bdelta}|^2]-|\mathbbm{E}[G_{\bgamma\bdelta}]|^2 \nonumber \\
     &= \sum_{\substack{\vr,\vs \\ \balpha,\bbeta}} \chi_{\balpha\bbeta} |G_{\bgamma\bdelta}(\vr,\vs)|^2 F_{\balpha\bbeta}(\vr,\vs) - |\chi_{\bgamma\bdelta}|^2 \nonumber \\
     &\leq \sum_{\substack{\vr,\vs \\ \balpha,\bbeta}} \chi_{\balpha\bbeta} |G_{\bgamma\bdelta}(\vr,\vs)|^2 F_{\balpha\bbeta}(\vr,\vs).
     \label{var1}
\end{align}
Making use of the single-qubit structure of the problem allows to write Eq.~\eqref{var1} as
\begin{equation} \label{pseudo_var}
   \sum_{\balpha,\bbeta} \chi_{\balpha\bbeta} \prod_{i = 1}^n \left(\sum_{r_i,s_i} |\ggd|^2 \fab\right).
\end{equation}
We now exploit the $d$-Pauli-weight channel assumption. Since $\alpha_i,\beta_i,\gamma_i$ or $\delta_i$ can differ from the identity on a fixed number of qubits (given by $4d$), we can split the above product in the following way:
\begin{align}\label{intuitive}
    \prod_{i = 1}^{n-4d} \left(\sum_{r_i,s_i} |g_{00}(r_i,s_i)|^2 f_{00}(r_i,s_i) \right) \times \nonumber \\
    \prod_{j = 1}^{4d} \left(\sum_{r_i,s_i} |\ggd|^2 \fab \right),
\end{align}
where we have assumed without loss of generality that on the first $n-4d$ qubits $\alpha_i,\beta_i,\gamma_i$ or $\delta_i$ are the identity. We can see that in order for this product (and therefore the variance) to not scale exponentially in system size it is sufficient to ask for
\begin{equation}\label{goo2}
    \sum_{r_i,s_i} |g_{00}(r_i,s_i)|^2 f_{00}(r_i,s_i) \leq 1.
\end{equation}
Enforcing this condition translates into the following quadratic minimization problem:
\begin{equation}\label{min}
    \min_{\vec{x}}\left\{\sum_{r_i,s_i}\left|g^{\mathrm{sh}}_{00} + \vec{x}^{\mathrm{T}}\mathbf{K}\right|^2\hspace{-0.3em}(r_i,s_i)f_{00}(r_i,s_i)\right\},
\end{equation}
where we have fixed the initial choice of $\vg$ to be the one in Eq.~\eqref{g_shh}.
Quadratic problems admit a well-known analytic solution, leading in this case to
\begin{equation} \label{our_g}
    g^{\mathrm{min}}_{00} (r_i,s_i) = \begin{cases}
        -\frac{7}{2} \quad \text{for } r_i = \bar{s}_i\\
        1 \quad \text{otherwise}
    \end{cases}   
\end{equation}
where with $r_i = \bar{s}_i$ we mean that $r_i$ and $s_i$ are in the same Pauli basis, but different eigenstates (e.g. $\ket{r_i} = \ket{+}$ and $\ket{s_i} = \ket{-}$). 
By direct inspection one can see that $g^{\mathrm{min}}_{00}$ saturates the inequality in Eq.~\eqref{goo2}. To obtain a better scaling of the sample complexity it is reasonable to apply the same minimization procedure defined in Eq.~\eqref{min} to all the entries $\gamma_i,\delta_i$,
\begin{equation}\label{total_min}
    \min_{\vec{x}}\left\{\sum_{r_i,s_i}\left|g^{\mathrm{sh}}_{\gamma_i\delta_i} + \vec{x}^{\mathrm{T}}\mathbf{K}\right|^2\hspace{-0.3em}(r_i,s_i)f_{00}(r_i,s_i)\right\}.
\end{equation}
We then define $\vg^{\mathrm{min}}$ as the matrix that results from the above minimization of all entries. We refer to Appendix~\ref{sample_complexity} for an explicit expression of such function.
Now, using $\vg^{\mathrm{min}}$, we can prove that	
\begin{equation} \label{var_bound}
    \text{Var}[G^{\mathrm{min}}_{\bgamma\bdelta}] \leq C^{4d},
\end{equation}
with $C$ being the smallest constant such that $\sum_{r_i,s_i} |g^{\mathrm{min}}_{\gamma_i\delta_i}(r_is_i)|^2 \fab \leq C$ for every $\alpha_i,\beta_i,\gamma_i,\delta_i$ and $\bG^{\mathrm{min}}$ being the product of $\vg^{\min}$ over all the qubits.  This last equation implies that, making use of the median of means estimator (MoM)~\cite{aaronson2018shadow}, the sample complexity $M$ to estimate with probability $1-\delta$ one entry of $\bchi$ with precision $\epsilon$ is
\begin{equation}\label{mom} 
    M = \map{O}\left(\frac{e^d}{\epsilon^d}\log\left(\frac{1}{\delta}\right)\right),
\end{equation}
which is constant in system size. If instead we wish to estimate the $\ell_2$-distance of the channel resulting from the estimated $\chi_{\balpha\bbeta}$ and the actual noise component $\map{E}$ we need

\begin{align}\label{global_scaling}
    M = \scalingeq,
\end{align}
which is now logarithmic in system size. Finally the scaling for the reconstruction in diamond distance is
\begin{align}\label{global_scaling_diamond}
    M = \scalingdiamond.
\end{align}
 . \\
It is interesting to note that, to obtain the bounds on the variance, it is sufficient to assume that the process is trace nonincreasing (as shown in Lemma~\ref{trace_noninc}). Consequently, our protocol can be readily applied to trace-nonincreasing channels, which describe, for instance, physical processes such as leakage~\cite{wood2018leakage}. \\
From now on when talking about $\vg$ we will mean the function resulting from the minimization procedure: $\vg^{\mathrm{min}}$.
\section{Classical postprocessing}\label{postprocessing}
The tomography protocol described above is efficient in the case of low-degree channels. Yet, what one has direct experimental access to is the noisy implementation $\map{C}$, which includes the unitary-layer contribution. Even in the case of low noise, such a map is far from being a low-degree channel, while the noisy part $\map{E} = \map{C}\circ\map{U}^{\dagger}$ is.
However, implementing $\map{U}^\dagger$ in the quantum device would create a new source of error and would not allow for the correct estimate of the noisy part $\map{E}$. 

We thus seek to implement the inverse gate directly in classical postprocessing, essentially by adapting the function $\bG$ to undo the action of $\map{U}$. This process is not guaranteed to preserve the efficiency of the tomography protocol. Yet, we find that the protocol remains scalable (with just a polynomial overhead) for the case of layers containing a polylogarithmic number of non-Clifford gates. \par
Let us explain the postprocessing protocol in detail: first, recall that
we want to reconstruct the process matrix of the noise component from Eq.~\eqref{G}, namely,
\begin{equation}
     \chi_{\balpha\bbeta}^{\map{E}} = \frac{1}{18^n} \sum_{\vr,\vs} \bra{\vr} \map{C}\circ \mapd{U}(\ketbra{\vs}{\vs}) )\ket{\vr}G_{\balpha\bbeta}(\vr,\vs).
\end{equation}
However, the randomized measurement data is sampled according to the noisy implementation of $\map{U}$, that is,
\begin{equation}\label{eq:pc}
    p_{\map{C}}(\vr,\vs) = \frac{1}{18^n} \bra{\vr} \map{C}(\ketbra{\vs}{\vs})\ket{\vr}.
\end{equation}
With simple manipulation we can rewrite
\begin{align}
    \chi_{\balpha\bbeta}^{\map{E}}
     &= \frac{1}{18^n} \sum_{\vec{r},\vec{s'}} \bra{\vec{r}} \map{C}(|\vec{s'}\rangle\langle\vec{s'}|)\ket{\vec{r}}G^{\map{U}}_{\balpha\bbeta}(\vec{r},\vec{s'})\nonumber \\
     & = \mathbb{E}_{\sim\map{C}}[G^{\map{U}}_{\balpha\bbeta}(\vec{r},\vec{s'})], \label{expectation_rot}
\end{align}
where we use the subscript $\sim\map{C}$ to indicate that the expectation value is taken with respect to the probability distribution generated by $\map{C}$. Also, we have explicated the action of $\mapd{U}$ on its components: $\mapd{U}(\ketbra{\vs}) = \sum_{\vec{s'}} u_{\vs,\vec{s'}} |\vec{s'}\rangle\langle\vec{s'}|$ and defined the rotated $\bG$ function,
\begin{align} \label{rot_g}
    G_{\balpha\bbeta}^{\map{U}}(\vec{r},\vec{s'}) &= \sum_{\vec{s}} u_{\vs,\vec{s'}}G_{\balpha\bbeta}(\vr,\vs).
\end{align}
In order to have efficient sample complexity we need the function above to have variance bounded by poly$(n)$ on the distribution~\eqref{eq:pc},
\begin{align} \label{var_U}
    \text{Var}&[G_{\bgamma\bdelta}^{\map{U}}] = \mathbbm{E}_{\sim\map{C}}[|G_{\bgamma\bdelta}^{\map{U}}|^2]-|\mathbbm{E}_{\sim\map{C}}[G_{\bgamma\bdelta}^{\map{U}}]|^2 \\ \label{here}
     &\leq\frac{1}{18^n}  \sum_{\vr,\vs} |G_{\bgamma\bdelta}^{\map{U}}(\vr,\vs)|^2 \bra{\vr}\map{C}(\ketbra{\vs})\ket{\vr} 
\end{align}
We now invert Eq.~\eqref{E} to write $\map{C} = \map{E}\circ\map{U}$, that is, the noisy implementation can be modeled by the noiseless $\map{U}$, followed by the error channel.
Then, Eq.~\eqref{here} becomes
\begin{align} \label{pseudo_variance}
    \frac{1}{18^n} \sum_{\balpha\bbeta} \chi_{\balpha\bbeta}^{\map{E}} \sum_{\vr,\vs} |G_{\bgamma\bdelta}^{\map{U}}(\vr,\vs)|^2 \bra{\vr}\sigma_{\balpha}\map{U}(\ketbra{\vs})\sigma_{\bbeta}\ket{\vr}
\end{align}
Let us now estimate the scaling of this term for a few interesting scenarios.
\par
\textit{Unitary on polylogarithmic number of qubits.} Let us consider a unitary map $\map{U}$ that only acts on a set $Q$ of qubits of cardinality $q$. Then, we can write
\begin{equation}
    G_{\bgamma\bdelta}^{\map{U}}(\vr,\vs) = G_{\bgamma\bdelta}^{Q}\left(\prod_{i \in \bar{Q}} \ggd \right),
\end{equation}
where $G_{\bgamma\bdelta}^{Q}$ is the part of the function supported on $Q$. It follows that $\chi_{\balpha\bbeta}^{\map{C}}$ will differ from $\chi_{\balpha\bbeta}^{\map{E}}$ on at most $q$ qubits for every $\balpha,\bbeta$, having nonzero entries for Pauli strings that have weight at most $2d + q$. This leads to a favorable sample complexity, with variance scaling as
\begin{align}
    \text{Var}(G_{\bgamma\bdelta}^{\map{U}}(\vr,\vs)) \leq K^q C^{4d},
\end{align}
where $K$ is a constant. Therefore by taking $q=\map{O}(\text{poly}\log(n))$ we can see that, accepting a polynomial overhead in $n$, the sample complexity remains efficient,
\begin{equation} \label{polylog_scaling}
    M = \map{O}\left(n^p\frac{e^d}{\epsilon}\log\left(\frac{1}{\delta}\right)\right),
\end{equation}
where $p$ is the product of the exponents appearing in the polynomial dependence on $\log$ of the support of such gates.  For a detailed derivation we refer to Appendix~\ref{fixed_U}.
\par
\textit{Layer of two-qubit gates.} Take a layer of $n/2$ unitaries $\map{U}_{\bj}$, acting each on a pair of qubits $\bj = (j_0,j_1)$. 
\begin{equation}\label{U_prod}
    \map{U} = \map{U}_{\textbf{0}} \otimes \map{U}_{\textbf{1}} \otimes \cdots \map{U}_{\textbf{n/2}},
\end{equation}
with no restriction on the connectivity of the device. By exploiting the tensor product structure,  $G_{\balpha\bbeta}^{\map{U}}$ becomes a product of $n/2$ two-qubit functions,
\begin{align}
    G_{\balpha\bbeta}^{\map{U}}(\vec{r},\vs) &=  \sum_{\vec{s'}} u_{\vs,\vec{s'}}G_{\balpha\bbeta}(\vec{r},\vec{s'}) \\ \label{this_one}
    & = \prod_{\bj\in J} g^{\map{U}_{\bj}}_{\balpha_{\bj}\bbeta_{\bj}} (\br_{\bj},\bs_{\bj}),
\end{align}
where we defined the functions $g^{\map{U}_{\bj}}_{\balpha_{\bj}\bbeta_{\bj}}(\br_\bj,\bs_\bj) = \sum_{\bs_{\bj}'}u_{\bs_{\bj}'\bs_{\bj}} g_{\alpha_{j_0}\beta_{j_0}}(r_{j_0},s_{j_0}')g_{\alpha_{j_1}\beta_{j_1}}(r_{j_1},s_{j_1}')$ acting on the pair of qubits $(j_0,j_1)$.
Since we assume $\map{E}$ to be a few-body channel, the relevant contribution to Eq.~\eqref{pseudo_variance} comes for terms with $\alpha_i,\beta_i,\gamma_i$ and $\delta_i$ equal to the identity. In fact, given the above product structure of $G_{\balpha\bbeta}^{\map{U}}(\vec{r},\vs)$, we can split the contributions similarly to what done in Eq.~\eqref{intuitive}. Then we can see that the exponential scaling can be avoided if the terms corresponding to all indexes equal to identity are smaller or equal to $1$. In Appendix~\ref{layer} we show that bounding such contributions results in the following minimization problem:
\begin{equation}\label{min_u}
    \min_{\vec{x}}\sum_{\br_\bj,\bs_\bj}|g^{\map{U}_{\bj}}_{00} + \vec{x}^{\mathrm{T}}\mathbf{{K}}|^2 p_{\map{U}_\bj}(\br_\bj,\bs_\bj),
\end{equation}
where $ p_{\map{U}_\bj}(\br_\bj,\bs_\bj) = \frac{1}{18^2} \bra{\br_{\bj}}\map{U}_\bj(\ketbra{\bs_{\bj}})\ket{\br_{\bj}}$. Notice that now, since we are considering a two-qubit space, $\vec{x}$ encodes $1040$ free parameters. \par
Solving the corresponding quadratic problem allows us to find that Clifford gates always correspond to a minimum of $1$. In Appendix~\ref{post_sample} we show that this leads to the variance scaling as
\begin{equation}
    \text{Var}(G_{\bgamma\bdelta}) \leq n{C'}^{4d},
\end{equation}
That is, one can estimate with probability $1-\delta$ one entry of $\bchi$ with precision $\epsilon$ via the relation
\begin{equation}\label{chi_avg}
\hat{\chi}_{\balpha\bbeta}^{\mathcal{E}} = \frac{1}{M}\sum_{k=1}^M \boldsymbol{G}_{\balpha\bbeta}^{\mathcal{U}} (\vec{r}^{(k)}, \vec{s}^{(k)}
  )
\end{equation}
with sample complexity scaling as
\begin{equation}\label{mom_pp}
    M = \map{O}\left(n\frac{e^d}{\epsilon}\log\left(\frac{1}{\delta}\right)\right).
\end{equation} 
Notice that now the sample complexity required to estimate the channel in $\ell_2$-norm is multiplied by a linear factor in system size,
\begin{equation}
    M = \scalingeqpp.
\end{equation}
The argument carries over to the diamond distance as well; however, because the scaling without postprocessing already contains a factor of $n^{\map{O}(d)}$, the scaling for reconstruction in diamond norm remains unchanged.
Despite the bound being polynomial in system size, one can find numerical evidence that in many relevant cases the scaling of Eq.~\eqref{mom_pp} is actually constant, similar to what found for the case without postprocessing (see Appendix~\ref{post_num} for more details). We also confirmed numerically that the postprocessing scheme for non-Clifford gates leads to an exponential sample complexity (see Appendix~\ref{post_num}). The code to generate $\vg^{\map{U}_{\textbf{j}}}$ for any one or two-qubit unitary is publicly available in Ref.~\cite{git}, alongside the precomputed functions for \textit{CNOT} and \textit{iSWAP} gates.
\par
Combining together the discussion of the two subcases, we find that our  postprocessing tomography protocol shows at most polynomial scaling for any layer of two-qubits gates with a polylogarithmic number of non-Clifford gates. In fact,  for the reconstruction in $\ell_2$-norm, we get a final sample complexity of
\begin{equation}
    M = e^{{\mathcal{O}}(d \log(1/\epsilon))} \cdot \map{O}\left(n^{p+1}\log\left(\frac{n}{\delta}\right)\right).
\end{equation}
Finally the scaling in diamond distance reads
\begin{equation}
    M = n^{\map{O}(d)+p}e^{{\mathcal{O}}(\log(1/\epsilon))} \cdot \map{O}\left(\log\left(\frac{1}{\delta}\right)\right).
\end{equation}

\section{Numerical analysis}\label{numerics}
To showcase the scalability our method, we simulated numerically the action of a known error channel and reconstructed its low-degree process matrix by means of our tomography protocol. We focus our numerically analysis on two key aspects: first we show the applicability of our model by computing numerically the variance of $\bG$ for a correlated, coherent, low-degree error channel. Then we simulate the sampling procedure in the case of a single-qubit tensor product error channel.\par
For the first task we chose an error channel $\map{E}(\rho) = \sum_{k=1}^n E_k \rho E_k^{\dagger}$ with Kraus operators of the form

\begin{figure}[H]
\centering
\includegraphics[width=\linewidth]{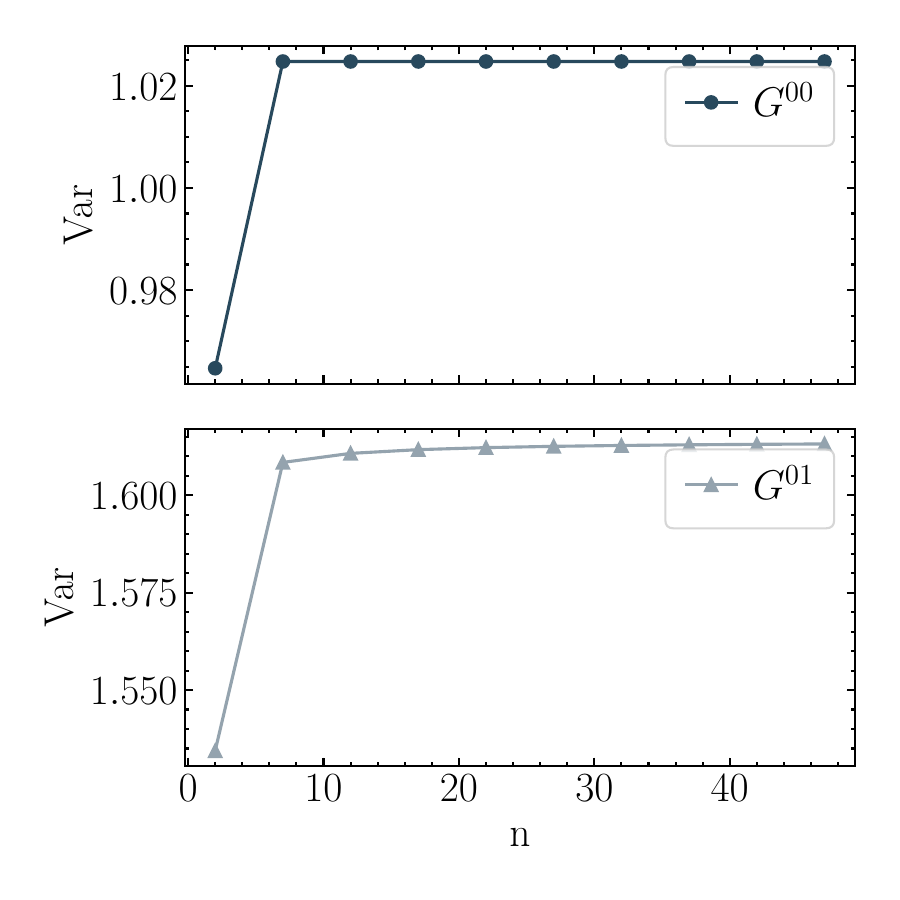}
\caption{Numerical computation of the variance of some entries of $\bG$ for an error channel with Kraus operators of the form~\eqref{single_qubit}. The computation is carried for the entries corresponding to $\balpha = \bbeta = (0,\ldots,0)$ (top panel) and $\balpha = (0,\ldots,0)$, $\bbeta = (x,0,\ldots,0)$ (bottom panel).}
\label{fig:var_M}
\end{figure}

\begin{align} \label{single_qubit}
    E_k = p\left(\Id + i\varepsilon\sum_{m=1}^nM_{km}\sigma_x^{(m)}\right),
\end{align}
where with $\sigma_x^{(m)}$ we mean the Pauli $\sigma_x$ acting on qubit $m$. We can enforce the normalization $\sum_k E_k^{\dagger}E_k = \Id$ by choosing $\boldsymbol{M}$ to be a normal matrix and $p = 1/(n(1+\varepsilon^2))$. We make the explicit choice $\boldsymbol{M} = \exp(-2/n\boldsymbol{A})$ with $\boldsymbol{A}$ being an anti-symmetric, tridiangonal matrix with period boundary conditions. This choice encapsulates the idea of correlated, coherent errors with magnitude decaying exponentially over an interaction length of $n/2$. Results are reported in Fig.~\ref{fig:var_M} for the entries of $\bG$ corresponding to $(\alpha_0,\beta_0) = (0,0)$ and $(0,x)$ and $0$ otherwise. The numerical analysis shows that the variance saturates to a constant.\par
For the simulation of the sampling procedure, we chose a 1D chain of qubits and the error channel as a tensor product of single-qubit channels of the form
\begin{dmath}\label{err_ch_sq}
     \map{E}_l (\cdot) = p_l\sigma_z e^{-i\gamma_l\sigma_x}(\cdot)e^{i\gamma_l\sigma_x}\sigma_z + \\
     (1-p_l)e^{-i\gamma_l\sigma_x}(\cdot)e^{i\gamma_l\sigma_x},
\end{dmath}
where the parameters $p$ and $\gamma$ decay exponentially with respect to the distance $l$ of the qubit to the center,
\begin{align} \label{p_l}
   p_l = p_0 \exp(-l), \quad
    \gamma_l = \gamma_0 \exp(-l).
\end{align}
We have chosen $p_0 = \gamma_0 = 0.1$.  It is important to notice that while this channel does not strictly respect the low-degree ansatz it can be approximated by one, well enough for our protocol to be efficient. For a detailed analysis of the error committed in such approximation for similar cases refer to Appendix~\ref{phys}. We run the simulation of the reconstruction of some entries of $\bchi$. More specifically we computed the number of samples needed to reach a fixed threshold of error $\epsilon$ for different system size, reaching up to $100$ qubits. \par
\begin{figure*}[t!]
\begin{subfigure}{0.49\linewidth}
\includegraphics[width=\linewidth]{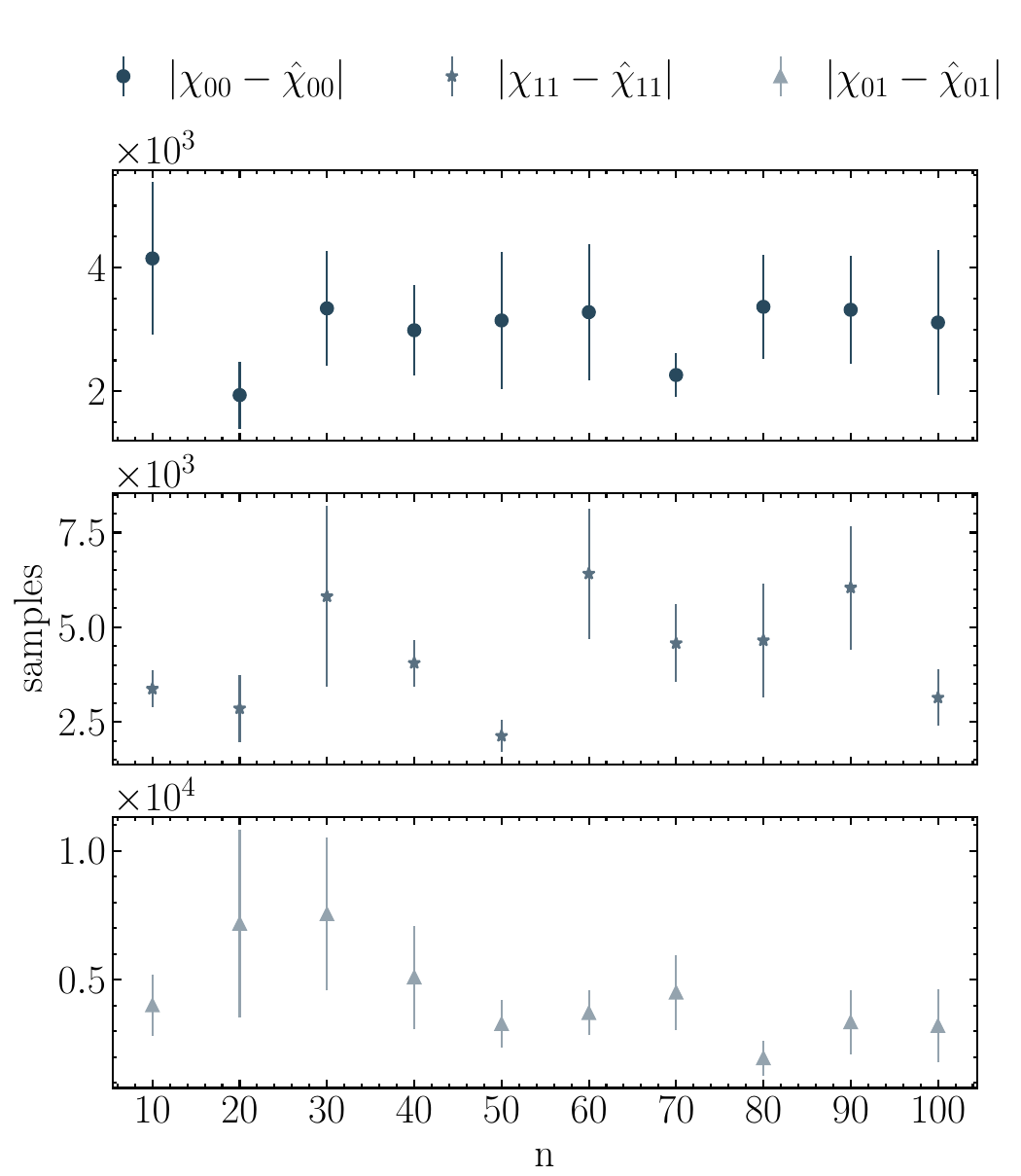}
\end{subfigure}
\hfill
\begin{subfigure}{0.49\linewidth}
\includegraphics[width=\linewidth]{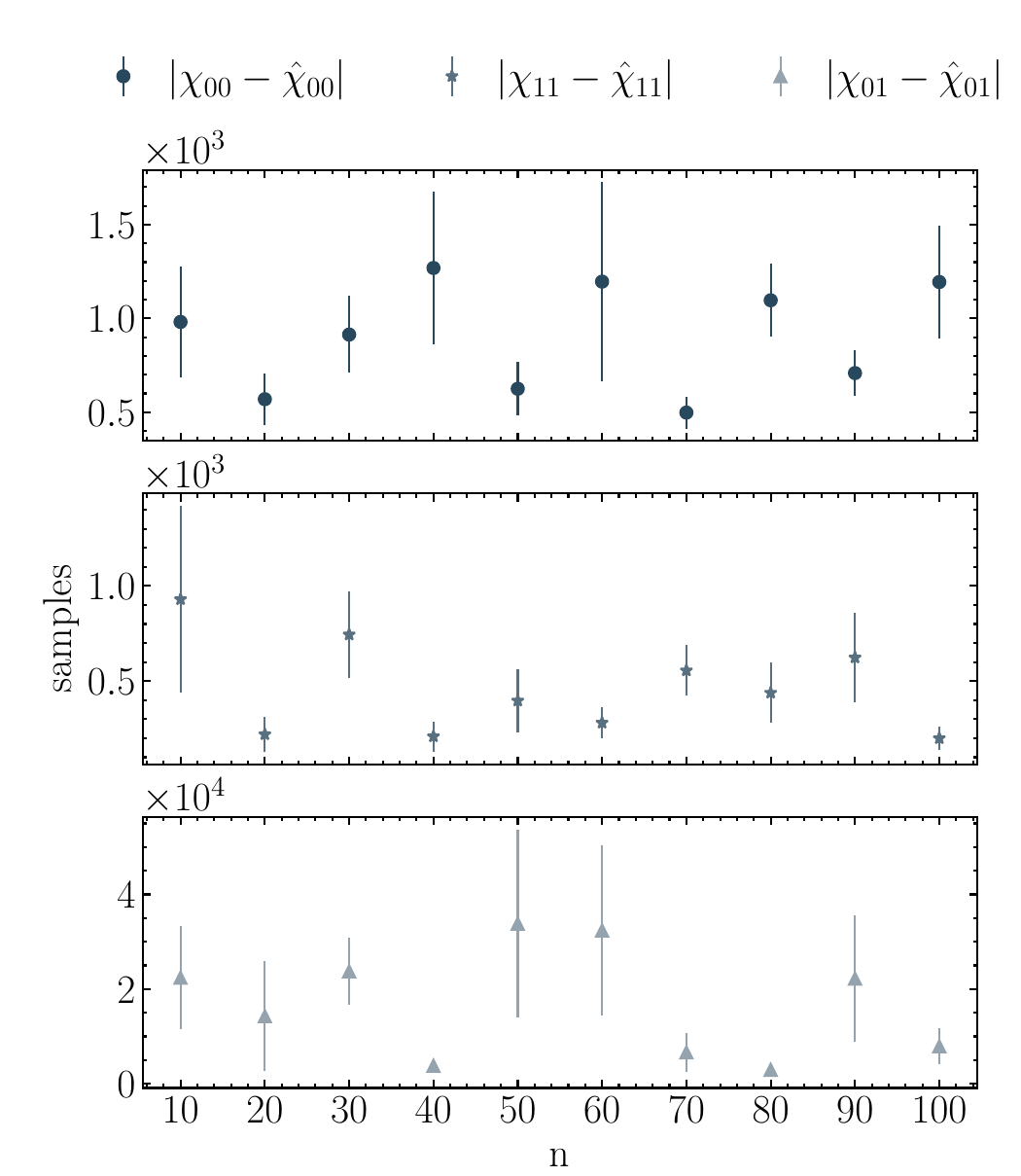}
\end{subfigure}%
\hfill
\begin{subfigure}{0.49\linewidth}
\centering
\includegraphics[width=0.8\linewidth]{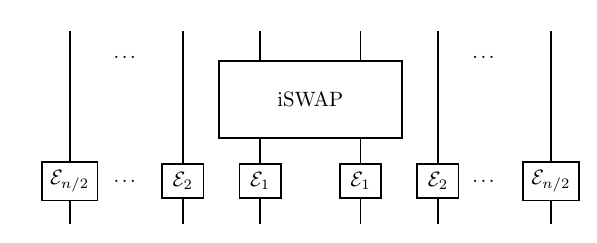}
\caption{}
\label{fig:single}
\end{subfigure}
\hfill
\begin{subfigure}{0.49\linewidth}
\includegraphics[width=\linewidth]{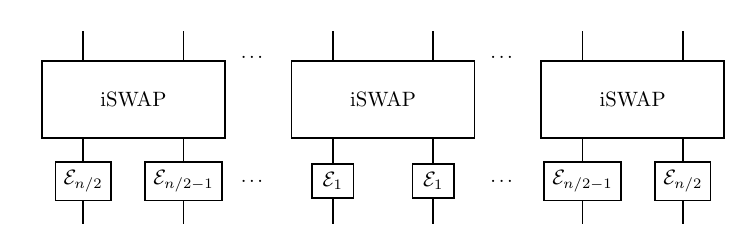}
\caption{}
\label{fig:layer}
\end{subfigure}%
\caption{\justifying \small{Number of samples needed to reach a precision of $0.05$ in the reconstruction of the entry corresponding to $\balpha = (\alpha_0,\ldots,0)$ and $\bbeta = (\beta_0,\ldots,0)$ for (from top to bottom) $(\alpha_0,\beta_0) = (0,0),(1,1),(0,1)$. We report the results for two systems: (a) single \textit{iSWAP} gate and (b) layer of \textit{iSWAP} gates. We considered the convergence to be reached if for $500$ consecutive samples $\abs{\hat{\chi}_{\balpha\bbeta}-\chi_{\balpha\bbeta}} < 0.05$, where with $\hat{\chi}_{\balpha\bbeta}$ we mean the process matrix resulting from sampling. Every point in the plots is the result of an average over 10 independent samplings and the error bar shows the standard deviation from the mean.  Below each plot is a representation of the corresponding circuit used in the numerics.}}
\label{fig:chi_00_thr}
\end{figure*}
\begin{figure}[H]
\includegraphics[width=0.95\linewidth]{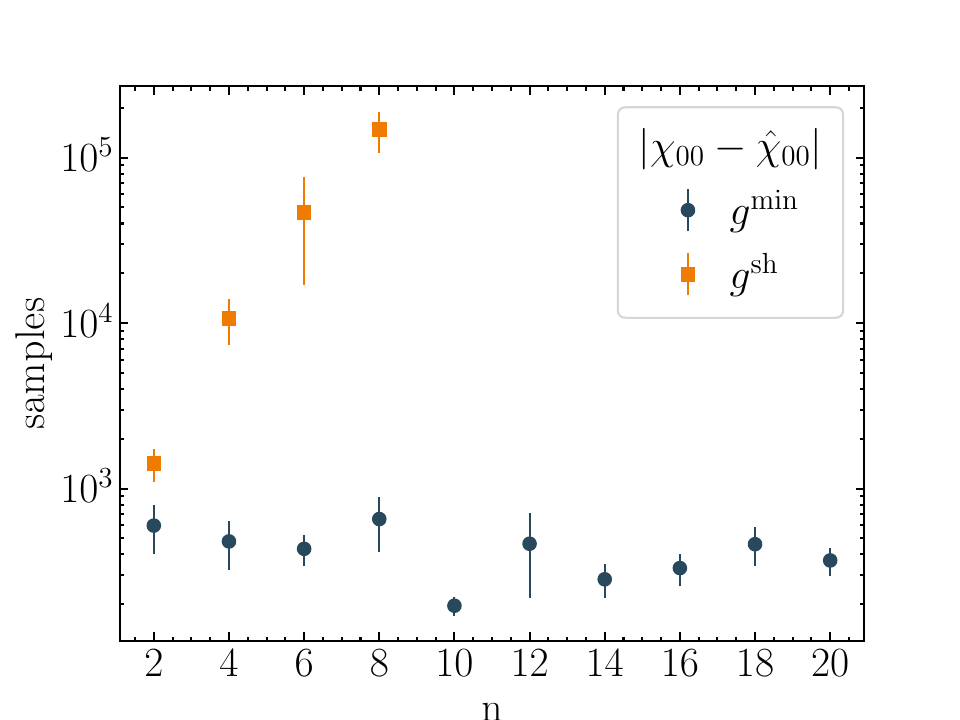}
    \caption{\small{Number of samples needed to reach a precision of $0.05$ in the reconstruction of the entry corresponding to $\balpha = \bbeta = (0,\ldots,0)$ for a layer of \textit{iSWAP} gates. We considered the convergence to be reached if for $100$ consecutive samples $\abs{\hat{\chi}_{\balpha\bbeta}-\chi_{\balpha\bbeta}} < 0.05$, where with $\hat{\chi}_{\balpha\bbeta}$ we mean the process matrix resulting from sampling. In blue the data obtained using the optimized function $\vg^{\min}$ and in orange the data from using $\vg^{\mathrm{sh}}$}. Every point in the plot is the result of an average over 10 independent samplings and the error bar shows the standard deviation from the mean.}
    \label{fig:comp}
\end{figure}

We studied two scenarios, depicted in Fig.~\ref{fig:comp}(\subref{fig:single}--\subref{fig:layer}). First, the application of a single \textit{iSWAP} gate in the center of the chain and then a full layer of nearest neighbors \textit{iSWAP} gates.
In the first case we do not make use of the rotated $\bG$ function and therefore the average converges to the process matrix $\bchi^{\map{C}}$ of the noisy implementation of the gate $\map{C} = \map{U}_{iSWAP} \circ \bigotimes_j \map{E}_j$. We made this choice since the noisy implementation of one single \textit{iSWAP} respects the ansatz of being low degree and therefore does not require the postprocessing procedure. In the second case, however, the contribution of the layer of gates generates a noisy channel $\map{C}$ with non-negligible higher-order Pauli components. Therefore, we make use of the rotated $\bG$ function defined in Eq.~\eqref{rot_g} to eliminate the action of the gates and optimized to minimize the variance (Eq.~\eqref{min_u}).
We find that its average value converges correctly to the process matrix of just the error channel $\bchi^{\map{E}}$.\par
We applied the procedure mentioned above for the reconstruction of three different entries of the process matrix. Figure~\ref{fig:chi_00_thr} reports the results for different values of $\balpha,\bbeta$. In the top panel $\balpha=\bbeta = (0,\ldots,0)$ which corresponds to the all-identity case and is related to the fidelity of the process. On the central panel we estimate the entry corresponding to $\alpha_0, \beta_0 = x$ and all other entries equal to zero, which encodes the probability of having a bit flip error on the first qubit. We then turned to estimating an off-diagonal element and chose the entry corresponding to $\alpha_0 = 0, \beta_0=x$ and all other entries equal to zero (bottom panel). Results reported show that for both cases we are able to reach the same error with a constant number of samples. 

In Fig.~\ref{fig:comp} we show the difference in the scaling of the number of samples needed to reach an error of $\epsilon = 0.05$, using two different $\vg$ functions. In blue we can see that using the optimized functions shows a constant scaling while in orange we see that the scaling is exponential if we simply use the function defined in Eq.~\eqref{rot_g} without minimizing the variance. While we are aware that shadow-tomography is not the most efficient method to compute the fidelity of a process (i.e. $\chi_{00}$), we report the comparison as a proof of principle that exponential advantage over classical shadow is possible. 

\FloatBarrier

\section{Conclusion}\label{conclusion}
In this work, we demonstrated an efficient tomography protocol for quantum channels that conform to a low-degree ansatz, which corresponds to a perturbative regime in terms of the number of Pauli errors. This ansatz was first introduced in Ref.~\cite{learning_lowdegree} and has since been employed in the analysis of various tomography methods~\cite{learning_lowdegree, wadhwa2024agnostic, sarovar2020detecting}. Our protocol ensures logarithmic complexity in the system size and is experimentally practical, requiring only the preparation of product states in Pauli eigenbasis and measurements in the same basis.
In particular, we provide a complete reconstruction of the channel in logarithmic sample complexity, with guarantees in the $\ell_2$-norm and diamond norm of the channel while only making use of single-qubit state preparation and measurement.
We showcase how to apply the tomography protocol to characterize the noise affecting a layer of two-qubit gates in a quantum computer. 
A key advantage of our approach is that the inversion of the unitary gate layer, for many relevant cases, is handled entirely through classical postprocessing, avoiding the need to physically implement the inverse on quantum hardware. While only requiring at most polynomial overhead in sample complexity. This allows for a self-consistent tomography procedure. \par
A natural next step is to investigate how commonly used figures of merit—such as average gate fidelity and diamond norm error—translate into the process matrix representation within our ansatz. Understanding this relationship could yield deeper insight into the operational performance of the reconstructed channels.  
Another important avenue concerns the impact of SPAM errors on the protocol. Since these errors remain a dominant source of inaccuracy in current quantum platforms~\cite{yu2025efficient}, studying how they affect our method—and developing strategies to mitigate them—will be essential for advancing the robustness and applicability of this framework in real-world experiments. In this regard, it may be possible to adapt techniques that have been successfully employed within the shadow-tomography formalism, such as robust shadow estimation~\cite{chen2021robust} and probabilistic error cancellation~\cite{temme2017error,jnane2024quantum}, to our setting. These approaches could provide a path toward mitigating SPAM-induced biases without requiring significant changes to the protocol. However, the precise adaptation of these techniques to our protocol goes beyond the scope of the present work and is left for future investigation.

\section*{Acknowledgments}
M.C. thanks Patrick Emonts, Giacomo Giudice, Erickson Tjoa, and Dominik Wild (in alphabetical order) for fruitful discussions. F.B. thanks Daniel Malz for insightful discussion during the initial stages of the projects.  We extend special thanks to Giacomo Giudice for valuable feedback on the manuscript. \\
We thank the anonymous Referee~B for insightful feedback that significantly improved this work, in particular by pointing out a way to reduce the scaling of the protocol from polynomial to logarithmic and to extend the postprocessing guarantee from a constant to a polylogarithmic number of non-Clifford gates. \\
M.C. and I.C. acknowledge support from the German Federal Ministry of Education and Research (BMBF) via the project FermiQP (No. 13N15889). The work at MPQ is partly funded by THEQUCO as part of the Munich Quantum Valley, which is supported by the Bavarian
state government with funds from the Hightech Agenda Bayern Plus. \\
F.B. acknowledges financial support from the ICSC – Centro Nazionale di Ricerca in High Performance Computing, Big Data and Quantum Computing, funded by European Union – NextGenerationEU and from European Union’s Horizon Europe research and innovation programme under the Marie Skłodowska-Curie Action for Project No. 101148556 (ENCHANT).  The
views and opinions expressed here are solely those of the authors and do not necessarily reflect those of the funding institutions. Neither of the funding institution can be held responsible for them.
\vspace{5cm}

\appendix
\begin{widetext}
\newpage 
\section{Physical models} \label{phys}
In this section we will provide explicit examples in which our ansatz approximates well an error channel. \par
In general to compute the distance in $\ell_2$-norm of the channel to its truncated version one would have to sum the modulus square of the elements with degree bigger than $d$. In fact for Parseval's identity for super operators we get~\cite{learning_lowdegree}
\begin{equation}\label{err_trunc}
    \norm{\map{E}-\map{E}_{d}}^2_2=\sum_{\balpha,\bbeta}\abs{\chi_{\balpha\bbeta}-\chi_{\balpha\bbeta}^d}^2 = \sum_{\balpha,\bbeta\notin I_d}\abs{\chi_{\balpha\bbeta}}^2,
\end{equation}
where we use the index $d$ to indicate the truncation at degree $d$. Since $\bchi$ is a positive semidefinite matrix this quantity can be bounded with the diagonal elements,
\begin{equation}\label{err_diag}
    \norm{\map{E}-\map{E}_{d}}^2_2 \leq \sum_{\balpha\notin I_d}\chi_{\balpha\balpha}\chi_{\bbeta\bbeta}.
\end{equation}
In the following we will compute this error in some interesting cases.

\subsection{Small single-qubit errors}
Assume that the noise acting on the quantum device can be represented as single-qubit noise acting on every qubit. For simplicity we will take the example of the a single-qubit bit flip, acting on qubit $i$ with probability $p$:
\begin{equation}
    \map{E}^{(i)} (\rho) = (1-p)\rho + p \sigma^{(i)}_x \rho_i \sigma^{(i)}_x.
\end{equation}
So the error channel acting on all $n$ qubits will be
\begin{equation}\label{sq_ch}
    \map{E}(\rho) = \bigotimes_{i = 1}^n \map{E}^{(i)} (\rho) = \sum_{k=0}^{n}  (1-p)^{n-k}p^{k} \sum_{i=0}^{\binom{n}{k}} P_i^{(k)} \rho P_i^{(k)},
\end{equation}
where $\{P_i^{(k)}\}_i$ are the $k-$degree Pauli $x$ strings, which are $\binom{n}{k}$. For small $p$ we can truncate the expression above up to a certain order. For instance if we choose to truncate to second order we will have
\begin{equation}\label{sq_trunc}
    \map{E}^{(i)} (\rho) = \left(1-np+\frac{ n(n-1)}{2}p^2\right)\rho + \left(p-(n-1)p^2\right)\sum_i \sigma^{(i)}_x \rho \sigma^{(i)}_x + p^2\sum_{i<j} \sigma^{(i)}_x\sigma^{(j)}_x \rho \sigma^{(j)}_x\sigma^{(i)}_x + \mathcal{O}(p^3),
\end{equation}
which respects the ansatz of Eq.~\eqref{channel_assump}. \par
To compute the error that we are making in the truncation we need to put together Eqs.~\eqref{err_trunc} and~\eqref{sq_ch}. 
\begin{equation}
    \norm{\map{E}-\map{E}_d}^2_2 = \sum_{k=d+1}^n \binom{n}{k} \left[(1-p)^{n-k}p^{k}\right]^2 \leq  \sum_{k=d+1}^n \binom{n}{k}
    (1-p^2)^{n-k}p^{2k} \leq \exp\left[-2n\left(\frac{d+1}{n}-p^2\right)\right],
\end{equation}
where we have used that $(1-p)^{2(n-k)}\leq(1-p)^{n-k}$ for every $p\leq1$ and then used Hoeffding's bound on the tail of the binomial distribution. If we want the error to be bounded by $\epsilon$ we need 
\begin{equation}
    p < \left[\frac{d+1}{n}-\frac{1}{2n}\log\left(\frac{1}{\epsilon}\right)\right]^{1/2}
\end{equation}
Therefore, for small enough error probability $p$ it is reasonable to expect low-degree Pauli channels to well represent the noise process.

\subsection{Spurious interactions}
Another common scenario is the presence of spurious interactions in a quantum device. In real life a quantum gate is implemented with an Hamiltonian evolution. 
Let us call $H_0$ the Hamiltonian that is implementing the desired gate. If such gate is a layer of two qubit gates then $H_0$ will contain terms acting nontrivially on the couples that are affected,
\begin{equation}
    H_0 = \sum_{\substack{\textbf{j}\in J \\ p,q}} g_{pq}^{j_0j_1} \sigma_{p}^{(j_0)}\sigma_{q}^{(j_1)}
\end{equation}
where $\{\textbf{j}\}_J$ is the set that includes the couples $(j_0,j_1)$ on which the two qubit gates are acting and $p,q$ are the indexes of the Pauli matrices $\{0,x,y,z\}$. All the other possible interactions are instead represented by the following Hamiltonian:
\begin{equation}
    H_{int} = \sum_{\substack{kl \\ pq}} c_{pq}^{kl}\sigma_{p}^{(k)}\sigma_{q}^{(l)},
\end{equation}
with $k,l$ running possibly on all qubits present in the system. So $H_{int}$ is a sum of local terms $h_{kl}=\sum_{pq}c_{pq}^{kl}\sigma_{p}^{(k)}\sigma_{q}^{(l)}$ . When the gates corresponding to these are not being implemented this interaction should not affect the system. However it is often the case that some spurious interactions are left in the system. When trying to implement the layer of gates, the system will then evolve according to
\begin{equation}
    \rho(t) = e^{-it(H_{g}+\alpha H_{int})}\rho \ e^{it(H_{g}+\alpha H_{int})}
\end{equation}
where $\alpha$ is a parameter that quantifies the amount of spurious interactions and is assumed to be small. Then the error channel defined in Eq.~\eqref{E} in this case would be
\begin{equation}
    \map{E}(\rho) = e^{-it(H_{g}+\alpha H_{int})}e^{itH_{g}}\rho e^{-itH_{g}}\ e^{it(H_{g}+\alpha H_{int})}.
\end{equation}
Now, we can write the derivative of the unitary evolution $U(t) =  e^{-it(H_{g}+\alpha H_{int})}e^{itH_{g}}$ as
\begin{equation}\label{der}
    \frac{d}{dt}U(t) =  e^{-it(H_{g}+\alpha H_{int})}(-i\alpha H_{int})e^{itH_{g}} = -i\alpha U(t)\tilde{H}_{int}(t),
\end{equation}
where we have defined $\tilde{H}_{int}(t) = e^{-itH_{g}} H_{int}e^{itH_{g}}$. Importantly we can notice that the Hamiltonian so defined only contains terms that act nontrivially on at most four qubits. We can see from Eq.~\eqref{der} that the unitary evolution can be written as
\begin{equation}
    U(t) = T_R\exp\left[-i\alpha\int_0^{t}ds\tilde{H}_{int}(s)\right],
\end{equation}
where with $T_R \exp$ we mean the inverse time ordered exponential. This can be expanded in powers of $\alpha$ as
\begin{align}
    U(t) &= \Id + i\alpha\int_0^{t}ds\tilde{H}_{int}(s)-\alpha^2\int_0^{t}ds_2\int_0^{s_2}ds_1\tilde{H}_{int}(s_2)\tilde{H}_{int}(s_1) + \map{O}(\alpha^3) 
\end{align}
The above truncated expression is again a low-degree evolution, where terms of order $k$ in $\alpha$ correspond to up to $4k$-weight operators. We can now call $S_k$ the term of order $k$ in the above sum and write
\begin{equation}
    \map{E}(\rho) = \left(\sum_{j=0}^\infty S_j\right)\rho\left(\sum_{j'=0}^\infty S^{\dagger}_{j'}\right) = \sum_{m=0}^{\infty}\sum_{k =0}^m S_{k}\rho S^{\dagger}_{m-k},
\end{equation}
where we have grouped the terms of the same order in $\alpha$. We can now write this in the process matrix representation, by expanding $S_j$ in the Pauli basis,
\begin{equation}
    \map{E}(\rho)  =  \sum_{m=0}^{\infty}\sum_{k =0}^m\left(\sum_\balpha c_\balpha^k P_\balpha\right)\rho \left(\sum_\bbeta (c_\bbeta^{m-k})^*P_\bbeta\right) = \sum_{\balpha,\bbeta}\sum_{m=0}^{\infty}\sum_{k =0}^m \chiab^{km} P_{\alpha}\rho P_{\beta},
\end{equation}
where we have defined $\chiab^{km}=c_\balpha^k(c_\bbeta^{m-k})^*$. To bound the difference in $\ell_2$-norm it, from Eq.~\eqref{err_trunc} we can write
\begin{equation}
    \norm{\map{E}-\map{E}_d}_2 = \sum_{\balpha,\bbeta \notin I_d} \abs{\sum_{m=0}^{\infty}\sum_{k =0}^m\chiab^{km}}^2 \leq \sum_{\balpha,\bbeta \notin I_d} \sum_{m=0}^{\infty}\sum_{k =0}^m\abs{\chiab^{km}}^2 
\end{equation}
We include in the sum all the terms for which $4m$ is bigger than $d$. In fact all the terms where one among $S_k$ and $S_{m-k}$ include contributions of order higher than $d/4$ are included in this set. Notice that those terms might contribute also to the lower degree part of the process matrix. Since those contributions are hard to isolate in general, we upper bound the sum by extending it to the set of all Paulis. Then we can write
\begin{equation}\label{long_ineq}
    \norm{\map{E}-\map{E}_d}_2 \leq \sum_{m = \lceil\frac{d}{4}\rceil}^\infty \sum_{k = 0}^m \sum_{\balpha,\bbeta} \abs{\chiab^{km}}^2 = \sum_{m = \lceil\frac{d}{4}\rceil}^\infty \sum_{k = 0}^m \sum_{\balpha,\bbeta} \abs{c_{\balpha}^{k}}^2\abs{(c_{\bbeta}^{m-k})^*}^2 =  \sum_{m = \lceil\frac{d}{4}\rceil}^\infty \sum_{k = 0}^m  \norm{S_k}_2\norm{S_{m-k}}_2.
\end{equation}
Let us now bound the $\ell_2$-norm of $S_k$,
\begin{equation}\label{spur_ineq}
\norm{S_k}_2 = \alpha^k\frac{\norm{\tilde{H}_{int}}_2^k t^k}{k!} \leq \frac{1}{k!}(\alpha htm)^k \leq \frac{1}{k!}(\alpha ht(2n)^2)^k,
\end{equation}
where we have used $\norm{\tilde{H}_{int}}_2 = \norm{H_{int}}_2 \leq \sum_{kl} \norm{h_{kl}}_2$, then we bounded the norm of each local term with a constant $h$. Finally we introduced $m$ as the number of terms present in $H_{int}$ which is smaller than $\binom{n}{2} \leq (2n)^2$ but could show smaller scaling (e.g. linear) for different device connectivity (e.g. nearest neighbors). Then from Eq.~\eqref{long_ineq} we get the bound
\begin{equation}
    \sum_{m = \lceil\frac{d}{4}\rceil}^\infty \sum_{k = 0}^m\left(\frac{1}{(m-k)!k!}\right)^2(\alpha ht(2n)^2)^{2m} \leq \sum_{m = \lceil\frac{d}{4}\rceil}^\infty (2\alpha ht(2n)^2)^{2m} = \frac{(2\alpha ht(2n)^2)^{2\lceil\frac{d}{4}\rceil}}{1-(2\alpha ht(2n)^2)^2} < 2(2\alpha ht(2n)^2)^{2\lceil\frac{d}{4}\rceil}.
\end{equation}
Here we have used $\sum_{k = 0}^m 1/(m-k)!k! \leq 2^m$ and the closed form solution of the geometric series, valid for $\alpha < 1/2ht(2n)^2$. Last inequality assumes that $2\alpha ht(2n)^2<1/2$. So to have error smaller than $\epsilon$ one would need
\begin{equation}
    \alpha < \frac{1}{2ht(2n)^2}\left(\frac{\epsilon}{2}\right)^{1/\left(2\lceil\frac{d}{4}\rceil\right)}.
\end{equation}
This condition is consistent with previous assumptions, provided that we can choose $\epsilon$ small enough. Therefore, as long as $\alpha$ is small enough with respect to system size, the error that we are making in truncating to a low-degree representation is negligible.

\section{Shadow-tomography function}\label{others}
An example of a function whose expectation values gives the desired observable (like $\ggd$ in our protocol~\eqref{inv}), is used in the well-known shadow-tomography protocol. In fact in such protocol one would link the quantity to estimate to a certain observable of the reconstructed state. In our case, the state to reconstruct is the Choi state of channel $\map{E}$,
\begin{equation}
    \hat{\chi}_{\bgamma\bdelta}= \Tr[O_{\bgamma\bdelta}\hat{\Lambda}_{\map{E}}]
\end{equation}
It is easy to see that $\chi_{\bgamma\bdelta} = \Tr[(\sigma_{\bgamma}\otimes\Id_n)\ketbra{\phi^+}^{\otimes n} (\sigma_{\bdelta} \otimes \Id_n)\Lambda_{\map{E}}]$, with $\ket{\phi^+} = (\ket{00}+\ket{11})/\sqrt{2}$. The reconstruction of the Choi state in shadow-tomography formalism is given by the average over the classical shadows of the input and output states as $\hat{\Lambda}_{\map{E}} = \sum_{\vec{r},\vec{s}} p(\vr,\vs)(\map{D}_{1/3}^{\otimes -n}(\ketbra{\vr})\otimes\map{D}_{1/3}^{\otimes -n }(\ketbra{\vs}))$, where we have defined $\map{D}_{\lambda}^{-1}$ as the inverse of the depolarizing channel with depolarizing probability $\lambda$. Putting it all together we have that in shadow-tomography formalism,
\begin{equation}
    \hat{\chi}_{\bgamma\bdelta} =\sum_{\vr,\vs}p(\vr,\vs) \Tr[(\sigma_{\bgamma}\otimes\Id _n)\ketbra{\phi^+}^{\otimes n} (\sigma_{\bdelta} \otimes \Id_n)(\map{D}_{1/3}^{\otimes -n}(\ketbra{\vr})\otimes\map{D}_{1/3}^{\otimes -n }(\ketbra{\vs}))],
\end{equation}
which translates to a single-qubit function that satisfies Eq.~\eqref{inv}
\begin{equation} \label{g_sh}
    g^{\mathrm{sh}}_{\gamma_i\delta_i}(r_i,s_i) = \Tr[(\sigma_{\gamma_i}\otimes\Id) \ketbra{\phi^+}( \sigma_{\delta_i} \otimes \Id)(\map{D}_{1/3}^{-1}(\ketbra{r_i})\otimes\map{D}_{1/3}^{-1}(\ketbra{s_i}))].
\end{equation}
As mentioned in the main text, the function obtained by the product of~\eqref{g_sh}, despite being well known in literature, has variance growing exponentially with system size and for this reason cannot be used in practice. In fact we can provide a lower bound for its variance by making use of the following property of the function:
\begin{equation}
    \sum_{r_i,s_i} |g_{\gamma_i\delta_i}^{\mathrm{sh}}(r_i,s_i)|^2 \fab = C_{\alpha_i\beta_i}^{\gamma_i\delta_i} \delta_{\alpha_i\beta_i} \quad \text{with } \  C_{\alpha_i\beta_i}^{\gamma_i\delta_i} \geq 11/8  \quad \forall \alpha_i,\beta_i,\gamma_i,\delta_i \, ,
\end{equation}
where we recall that $\fab = (1/18) \bra{r_i} \sigma_{\alpha_i} \ketbra{s_i}{s_i} \sigma_{\beta_i}\ket{r_i}$.
Then we can write the variance of $G_{\bgamma\bdelta}$ using $\vg^{\mathrm{sh}}$ as
\begin{align}
    \text{Var}(G^{\mathrm{sh}}_{\bgamma\bdelta}) &= \sum_{\balpha,\bbeta} \chi_{\balpha\bbeta} \prod_{i = 1}^n \left(\sum_{r_i,s_i} |g^{\mathrm{sh}}_{\gamma_i\delta_i}(r_i,s_i)|^2 \fab\right) - |\chi_{\bgamma\bdelta}|^2 \\
    &\geq \left(\frac{11}{8}\right)^n \sum_{\balpha,\bbeta} \chi_{\balpha\bbeta} \delta_{\balpha\bbeta} - |\chi_{\bgamma\bdelta}|^2 \\
    &\geq \left(\frac{11}{8}\right)^n - 1  
\end{align}
where we used the fact that the diagonal of $\chi$ sums to $1$ and the fact that for every entry of $\bchi$ we have $\abs{\chi_{\gamma\delta}}\leq1$. Notice that the above bound clearly scales exponentially with system size $n$.

\section{Proofs of sample complexity} \label{sample_complexity}
In this section we will prove all the statements regarding sample complexity for our tomography protocol presented in Sec.~\ref{rm}. \\
First let us prove under which conditions the variance of the function $\bG$ defined in~\eqref{G} remains constant with increasing system size. Such function can be expressed as a product of single-qubit functions that we call $\vg$ defined in~\eqref{inv}. There are two properties that we wish $\boldsymbol{g}$ to fulfill in order for the variance of $\bG$ to be constant,
\begin{enumerate}[label=(\roman*)]
    \item $\sum_{r_i,s_i} |g_{00}(r_i,s_i)|^2 f_{00}(r_i,s_i) \leq 1$;
    \item $\sum_{r_i,s_i} |\ggd|^2 \fab = C_{\alpha_i\beta_i}^{\gamma_i\delta_i} \delta_{\alpha_i\beta_i} \quad \forall \ \gamma_i,\delta_i$ for some constant factor $C_{\alpha_i\beta_i}^{\gamma_i\delta_i}$,
\end{enumerate}
where $\vf$ are the functions defined in Eq.~\eqref{f_ab}.
Let us now prove that these two conditions imply constant variance. 
The first thing to notice is that, given our ansatz, the number of elements for which $\alpha_i,\beta_i,\gamma_i$ or $\delta_i$ differ from the identity is constant and simply given by four times the Pauli weight of the problem: $4d$. So if $C$ is the biggest constant such that $ \sum_{r_i,s_i} |\ggd|^2 \fab \leq C$ for one among $\alpha_i,\beta_i,\gamma_i$ or $\delta_i$ different from zero, we have that
\begin{align} 
     \sum_{\balpha,\bbeta \in I_d} &\chi_{\balpha\bbeta} \prod_{i = 1}^n \left(\sum_{r_i,s_i} |\ggd|^2 \fab\right) = \\
     &\leq C^{4d}  \sum_{\balpha,\bbeta \in I_d} \chi_{\balpha\bbeta} \delta_{\balpha\bbeta} \\
     &= C^{4d} \Tr[\bchi] \leq C^{4d},
\end{align}
where in the first line we have specified the summation over the low-degree Paulis and used condition (ii). From line one to line two we have used the fact that $C_{00}^{00} \leq 1$ (condition (i)) and the limited number of elements in $I_d$ that differ from zero. Lastly we used the fact that $\bchi$ has trace equal or minor to one, due to trace non increasing property of the channel. Therefore if we were to find a function that satisfies conditions (i) and (ii) we would have
\begin{equation}\label{single_entry_var}
    \text{Var}[G_{\balpha\bbeta}] \leq C^{4d}.
\end{equation}
For the sake of completeness let us report here the full description of the function $\vg^{\min}$ obtained through the minimization procedure of Eq.~\eqref{total_min}, dropping the qubit index for clarity.
\begin{align} \label{our_full_g}
    g^{\mathrm{min}}_{00} (r,s) &= \begin{cases}
        -\frac{7}{2} \quad \text{for } r = \bar{s}\\
        1 \quad \text{otherwise} 
        \end{cases} \\
    g^{\mathrm{min}}_{\alpha\alpha} (r,s) &= \begin{cases}
        (-1)^{\sigma_{b(r)}\star \sigma_\alpha}\frac{9}{2} \quad \text{ for } r = \bar{s}\\
        0 \quad \text{otherwise}
        \end{cases} \\
    g^{\mathrm{min}}_{\alpha\beta} (r,s) &= \begin{cases}
        \left((-1)^{e(r)+ e(s)}\frac{9}{4}\right) (\delta_{b(r)\alpha}\delta_{b(s)\beta} +\delta_{b(r)\beta}\delta_{b(s)\alpha} ) \\ 
        (-1)^{e(s)}\frac{9}{2}\quad \text{ for }e(r) \neq e(s)\text{ and }b(r) = b(s) = [\sigma_{\alpha},\sigma_{\beta}] \\
        0 \quad \text{otherwise}
        \end{cases} \\
    g^{\mathrm{min}}_{0\alpha}(r,s) = (g^{\mathrm{min}}_{\alpha0})^* (r,s) &= \begin{cases}
        \varepsilon_{b(r)b(s)\alpha}(-1)^{e(r)+ e(s)} \frac{9}{4}i\quad \text{ for }b(r),b(s) \neq \alpha\\
        \left((-1)^{e(r)} \frac{9}{10}\delta_{b(r)\alpha} + (-1)^{e(s)} \frac{9}{10} \delta_{b(s)\alpha}\right)\left(\frac12\right)^{\delta_{b(r)\alpha}\delta_{b(s)\alpha}} \text{ otherwise}\\
        \end{cases} 
\end{align}
where we recall that with $r=\bar{s}$ we mean that $r$ and $s$ are in the Pauli same basis, but with opposite eigenstates. Then with $b(r)$ we indicate the basis of which $r$ is eigenvector and with $e(r) \in \{0,1\}$ the eigenvector. For instance $\ket{-}$ would have $b(-) = x$ and $e(-) = 1$. We have also introduced the operation $\star$ between two Pauli matrices that is so defined as
\begin{equation}
     \sigma_{\alpha}\star \sigma_{\beta}= \begin{cases}
        0 \quad [\sigma_{\alpha},\sigma_{\beta}] = 0\\
        1 \quad \text{otherwise} 
        \end{cases} \\
\end{equation}
Lastly with $\varepsilon_{ijk}$ we mean the Levi-Civita tensor. 
Then we can see from direct computation that this function satisfies conditions (i) and (ii). We will now derive the exact expression of the sample complexity with respect to the variance of the function $\bG^{\min}$, where we will drop the superscript. \\
Let us start from the well-known Chebyshev's inequality,
\begin{equation} 
    \text{Pr}(|\hat{\chi}_{\balpha\bbeta} - \chi_{\balpha\bbeta}| > \epsilon) \leq \frac{\text{Var}(G_{\balpha\bbeta})}{M \epsilon^2}.
\end{equation}
where $\hat{\chi}_{\balpha\bbeta}$ is the estimator for $\chi_{\balpha\bbeta}$ and $M$ is the total number of samples. In order to have that with probability $1-\delta$ the estimated value does not differ from $\chi_{\balpha\bbeta}$ more than $\epsilon$ we need
\begin{equation}
    M \sim \frac{\text{Var}(G_{\balpha\bbeta})}{\delta \epsilon^2}.
\end{equation}
This expression, however has a scaling of $1/\delta$ which is particularly bad. To avoid this we make use of the MoM estimator. \\
The MoM estimator works as follows: we divide our samples into $K$ subsamples of dimension $b$, i.e. $M=Kb$. For each subsample we compute the mean of the samples, which leads to $K$ estimators of $\hat{\chi}_{\balpha\bbeta}$,
\begin{equation}
    \hat{\chi}_{\balpha\bbeta}^{(1)}, \ldots, \hat{\chi}_{\balpha\bbeta}^{(K)}.
\end{equation}
The median of means estimator is then taken to be the median of these estimators,
\begin{equation}
    \hat{\chi}_{\balpha\bbeta}^{MoM} = \text{median}\{\hat{\chi}_{\balpha\bbeta}^{(1)}, \ldots, \hat{\chi}_{\balpha\bbeta}^{(K)}\}.
\end{equation}
If we now fix the size of the samples to be $b = 34\text{Var}(G_{\balpha\bbeta})/\epsilon^2 $ we get the bound
\begin{equation}\label{union_bound}
    \text{Pr}(|\hat{\chi}_{\balpha\bbeta}^{MoM} - \chi_{\balpha\bbeta}| > \epsilon) \leq 2e^{-\frac{K}{2}}.
\end{equation}
In order to have an error smaller than $\epsilon$ with probability smaller than $1-\delta$ we need $K = 2\log(\frac{2}{\delta})$. Putting everything together we get
\begin{equation}\label{sample_complexity_variance}
    M \sim \log\left(\frac{1}{\delta}\right) \frac{\text{Var}(G_{\balpha\bbeta})}{\epsilon^2} .
\end{equation}
Therefore we can now state
\begin{lemma}\label{scaling}
Let $\balpha, \bbeta \in \{0,1,2,3\}^{n}$. Then $\chiab$ can be estimated with error $\epsilon$ and probability $1-\delta$ using only single-qubit state preparation and measurement with
\begin{equation}
  \map{O}\!\left(\frac{\text{Var}(G_{\balpha\bbeta})}{\epsilon^{2}} \log \frac{1}{\delta}\right) = \map{O}\!\left(\frac{C^{d}}{\epsilon^{2}} \log \frac{1}{\delta}\right)
\end{equation}
queries to $\map{E}$. 
\end{lemma}
\begin{proof}
    Putting together Eq.~\eqref{single_entry_var} and~\eqref{sample_complexity_variance} we obtained the desired result.
\end{proof}
We now wish to estimate the corresponding channel up to error $\epsilon$ in the $\ell_2$-norm. To this end we adapt the proof of Theorem~1 from Ref.~\cite{learning_lowdegree} to our measurement protocol. A central technical ingredient in that work is a Bohnenblust--Hille type inequality for maps measured in the $S_1\to S_\infty$ operator norm; we recall the relevant definitions here for the reader's convenience. \\
For a matrix $M$ we denote by $\|M\|_{S_1}$ its Schatten–1 (trace) norm and by $\|M\|_{S_\infty}$ its Schatten–$\infty$ (operator) norm. For a linear superoperator $\Phi:\mathbb{C}^{d\times d}\to\mathbb{C}^{d\times d}$ we define the induced norm from $S_1$ to $S_\infty$ as
\begin{equation}\label{def:S1Sinf}
  \|\Phi\|_{S_1\to S_\infty}
  := \sup_{M\neq 0}\frac{\|\Phi(M)\|_{S_\infty}}{\|M\|_{S_1}}.
\end{equation}
We will also use the induced $S_1\to S_1$ norm
\(
  \|\Phi\|_{S_1\to S_1} := \sup_{M\neq 0}\|\Phi(M)\|_{S_1}/\|M\|_{S_1},
\)
and the standard fact that for any matrix $A$ one has $\|A\|_{S_\infty}\le\|A\|_{S_1}$. \\
For convenience we also report the main technical statement from Ref.~\cite{learning_lowdegree}, which we generalize to trace nonincreasing maps.
\begin{lemma}[Bohnenblust--Hille inequality for trace nonincreasing maps]\label{trace_noninc}
    Let $\map{E}$ be a completely positive, trace nonincreasing map of degree at most $d$, with associated process matrix $\bchi$. Then there exists a constant $K$ such that
    \begin{equation}
          \norm{\bchi}_{\frac{2d}{d+1}} \;\leq\; K^{d}.
    \end{equation}
\end{lemma}
\begin{proof}
     By Ref.~\cite{learning_lowdegree} Theorem~12 we have the general bound $\norm{\bchi}_{\frac{2d}{d+1}} \;\leq\; K^{d} \norm{\map{E}}_{S_1\to S_{\infty}}$. We are only left to prove that ${\norm{\map{E}}_{S_1\to S_{\infty}}\leq 1}$ for any completely positive, trace-nonincreasing map $\map{E}$.
Now note that for any such map $\map{E}$ and any operator $M$,
\begin{equation}
    \norm{\map{E}(M)}_1 \le \norm{M}_1.
\end{equation}
This is a standard contractivity property of trace-nonincreasing CP maps. Therefore, $\norm{\map{E}}_{S_1 \to S_1} \le 1$. Finally, since
\begin{equation}
    \norm{\map{E}}_{S_1 \to S_\infty} \le \norm{\map{E}}_{S_1 \to S_1},
\end{equation}
we obtain $\norm{\map{E}}_{S_1 \to S_\infty} \le 1$. Plugging this back into the Bohnenblust--Hille inequality gives the desired bound.
\end{proof}
We are now ready to prove the theorem that provides the scaling of our tomography protocol.
\begin{theorem}\label{our_protocol_thm}
Let $\map{E}$ be an $n$-qubit degree-$d$ quantum channel. There is an algorithm that learns $\map{E}$ in $\ell_{2}$-distance with precision $\epsilon$ and failure probability $\delta$, using only single-qubit state preparation and measurements with
\begin{equation}
\scalingeq
\end{equation}
queries to $\map{E}$.
\end{theorem}
\begin{proof}
We first state the algorithm. Let $K$ be the constant in the statement of Lemma~\ref{trace_noninc} and $C$ that of Lemma~\ref{scaling}.
\begin{algorithm}[H]
\caption{Learning low-degree channels with single-qubit state preparation and measurements, via BH inequality}
\begin{algorithmic}[1]
\State \textbf{Input:} A quantum channel $\map{E}$ of degree at most $d$, error $\epsilon$, and failure probability $\delta$
\State Let $c = \epsilon^{4d+2} K^{-4d^{2}}$
\State Perform $M_{1} = \map{O}\!\left(\frac{C^d}{c^{2}} \log \frac{(4n)^d}{\delta}\right)$ randomized Pauli measurements of $\map{E}$ to estimate $(\chi_{\balpha,\balpha})_{\balpha}$
\State Let $(\chi'_{\balpha,\balpha})_{\balpha}$ be the associated empirical distribution
\State Perform $M_2 = \map{O}\!\left(\frac{C^d}{c^{2} \epsilon^{2}} \log\frac{1}{c^{2}\delta}\right)$ randomized Pauli measurements of $\map{E}$
\For{$\balpha,\bbeta \in X_{c} = \{\balpha : |\chi'_{\balpha,\balpha}| \geq c\}$}
  \State Approximate $\chi_{\balpha,\bbeta}$ with $\chi''_{\balpha,\bbeta}$ using Lemma~\ref{scaling}
\EndFor
\State \textbf{Output:} $\sum_{\balpha,\bbeta\in X_{c}} \chi''_{\balpha,\bbeta} P_{\balpha} (\cdot) P_{\bbeta}$
\end{algorithmic}
\end{algorithm}
Let $c > 0$ be determined later. In the first part of the algorithm we perform $M_{1} = \map{O}(C^d/c^{2} \log((4n)^d/\delta))$ randomized single-qubit Pauli measurements on $\map{E}$. We use the results to estimate the diagonal of the corresponding process matrix $\bchi$.   
Let $(\chi'_{\balpha,\balpha})_{\balpha\in\{0,1,2,3\}^{n}}$ be the empirical distribution after $M_{1}$ samples.
Consider the event,
\begin{equation}
E_1 = \{\,|\chi_{\balpha,\balpha} - \chi'_{\balpha,\balpha}| \leq c, \;\;\forall \balpha \in \{0,1,2,3\}^{n}\,\}.
\end{equation}
Then the complementary event is equivalent to
\begin{align}
    \bar{E}_1 = \bigcup_{\balpha} \{\,|\chi_{\balpha,\balpha} - \chi'_{\balpha,\balpha}| >  c\} := \bigcup_{\balpha}\bar{E}_1^{\balpha}
\end{align}
So for the union bound
\begin{equation}
    \Pr[\bar{E}_1] \leq \sum_{\balpha} \Pr[E_1^{\balpha}] \leq \delta,
\end{equation}
where we have used the fact that by Lemma~\ref{scaling}, taking $M_{1} = \map{O}(C^d/c^{2} \log((4n)^d/\delta))$ ensures that $\Pr[\bar{E}_1^{\balpha}] < \delta/(4n^d)$ for every $\balpha \in \{0,1,2,3\}^{n}$. Therefore we get that by taking $M_1$ samples we obtain $\Pr[E_1] \geq 1- \delta$. \\
Define $X_{c} = \{\balpha : |\chi'_{\balpha,\balpha}| \geq c\}$. Since $\sum_{\balpha\in X_{c}} \chi_{\balpha,\balpha} \leq 1$, we know
\begin{equation} \label{card}
|X_{c}| \leq c^{-1}.
\end{equation}
Assuming $E_1$ holds, for $\balpha \notin X_{c}$ we have
\begin{equation} \label{tr}
|\chi_{\balpha,\balpha}| \leq |\chi'_{\balpha,\balpha}| + \big||\chi_{\balpha,\balpha}| - |\chi'_{\balpha,\balpha}|\big| \leq 2c.
\end{equation}
In particular,
\begin{equation} \label{elem}
\balpha \notin X_{c} \;\;\Rightarrow\;\; |\chi_{\balpha,\bbeta}| \leq \sqrt{|\chi_{\balpha,\balpha}|\,|\chi_{\bbeta,\bbeta}|} \leq \sqrt{2c}, \quad \forall \bbeta,
\end{equation}
where the first inequality uses that $\bchi$ is positive semidefinite and the second uses~\eqref{tr} and $|\chi_{\bbeta,\bbeta}|\leq 1$. \\
In the second part of the algorithm, by Lemma~\ref{scaling} and with 
\begin{equation}
M_{2} = \map{O}\!\left(\frac{C^d}{c^{2}\epsilon^{2}} \log\frac{1}{c^{2}\delta}\right),
\end{equation}
we can find approximations $\chi''_{\balpha,\bbeta}$ of $\chi_{\balpha,\bbeta}$ for every $\balpha,\bbeta\in X_{c}$ such that
\begin{equation}
\sup_{(\balpha,\bbeta)\in X_{c}\times X_{c}} |\chi_{\balpha,\bbeta} - \chi''_{\balpha,\bbeta}| \leq c\epsilon
\end{equation}
with probability at least $1-\delta$. In fact let us consider the following event:
\begin{equation}
    E_2 = \Pr[|\chi_{\balpha,\bbeta} - \chi''_{\balpha,\bbeta}| \leq c\epsilon, \;\; \forall (\balpha,\bbeta)\in X_{c}\times X_{c}].
\end{equation}
Then its complement reads
\begin{equation}
    \Pr\left[\bigcup_{(\balpha,\bbeta)\in X_{c}\times X_{c}}|\chi_{\balpha,\bbeta} - \chi''_{\balpha,\bbeta}| > c\epsilon\right] \leq \sum_{(\balpha,\bbeta)\in X_{c}\times X_{c}} \Pr\left[|\chi_{\balpha,\bbeta} - \chi''_{\balpha,\bbeta}| > c\epsilon\right].
\end{equation}
Then by using $M_2 = \map{O}(K^d/c^2\epsilon^2 \log(1/c^2\delta))$ from Lemma~\ref{scaling} we get that
\begin{equation}
    \Pr[\bar{E}_2] \leq \frac{1}{c^2} c^2\delta = \delta,
\end{equation}
where we have used~\eqref{card}.
Let $\map{E}'' (\cdot) = \sum_{\balpha,\bbeta\in X_{c}} \chi''_{\balpha,\bbeta}\,\sigma_{\balpha}(\cdot)\sigma_{\bbeta}$. Then,
\begin{align}
\|\map{E} - \map{E}''\|_{2}^{2}
&= \sum_{\balpha,\bbeta\in X_{c}} |\chi_{\balpha,\bbeta} - \chi''_{\balpha,\bbeta}|^{2}
+ \sum_{\balpha\lor \bbeta\notin X_{c}} |\chi_{\balpha,\bbeta}|^{2} \\
&\leq \epsilon^{2} 
   + \sum_{\balpha \lor \bbeta \notin X_{c}} 
     \left|\chiab\right|^{\frac{2}{d+1}} 
     \left|\chiab\right|^{\frac{2d}{d+1}} \\
&\leq \epsilon^{2} 
   + (2c)^{\frac{1}{d+1}} 
     \left\lVert \map{E} \right\rVert_{\frac{2d}{d+1}}^{\frac{2d}{d+1}} \\
& \leq \epsilon^{2} + c^{1/(d+1)} K^{d}.
\end{align}
where in the equality we have used Parseval’s identity; in the first inequality we used Eq.~\eqref{card}, the learning guarantees of the second part of the algorithm and that $2 = 1/(d + 1/2) + 2d/(d + 1/2)$; in the second inequality we have used Eq.~\eqref{elem}; and in the third inequality we used the Bohnenblust-Hille inequality for trace nonincreasing maps (Lemma~\ref{trace_noninc}). \\
Hence choosing $c = \epsilon^{2d+2} K^{-d(d+1)}$ and simplifying the expression using $\log(1/\epsilon) \gtrsim d$  yield the desired result. 
\end{proof}
 
We will now adjust Theorem~\ref{our_protocol_thm} to reconstruct the channel with recovery guarantee in diamond distance. To do that we will make use of the following result.
\begin{lemma}\label{diamond_bound}
    Let $\map{E}$ and $\map{E}'$ be two quantum channels, with process matrix respectively $\bchi$ and $\bchi'$, then the following inequality holds
    \begin{equation}
        \norm{\map{E}-\map{E}'}_{\diamond} \leq \sum_{\balpha,\bbeta} |\chiab-\chiab'|
    \end{equation}
\end{lemma}
\begin{proof}
    Let us write the definition of diamond distance between two channels,
    \begin{equation}
        \norm{\map{E}-\map{E}'}_{\diamond} = \sup_{\rho}\norm{(\map{E}\otimes\Id)\rho -(\map{E}'\otimes\Id)\rho }_1,
    \end{equation}
    where the maximization is done over all density matrices $\rho$ of dimension $2^{2n}$. 
    Then the following inequalities hold
    \begin{align}
        \norm{\map{E}-\map{E}'}_{\diamond} & = \sup_{\rho}\norm{\sum_{\balpha,\bbeta} (\chiab-\chiab') (P_{\balpha}\otimes\Id)\rho (P_{\bbeta}\otimes\Id)}_1 \\
        &\leq \sum_{\balpha,\bbeta} |\chiab-\chiab'| \  \sup_{\rho}\norm{ (P_{\balpha}\otimes\Id)\rho (P_{\bbeta}\otimes\Id)}_1 \\
        &\leq \sum_{\balpha,\bbeta} |\chiab-\chiab'| \ \sup_{\rho}\norm{ (P_{\balpha}\otimes\Id)}_{\infty}\norm{\rho}_1\norm{(P_{\bbeta}\otimes\Id)}_{\infty},
    \end{align}
    where from line one to two we have used the triangular inequality and from line two to three we have used H\"older's inequality. Lastly, knowing that the infinity norm of $P_{\balpha}\otimes\Id$ and the one norm of $\rho$ are one we can finally write
    \begin{equation}
        \norm{\map{E}-\map{E}'}_{\diamond} \leq \sum_{\balpha,\bbeta} |\chiab-\chiab'|,
    \end{equation}
    which concludes the proof.
\end{proof}
We are now ready to state the algorithm for reconstruction in diamond distance.
\begin{theorem}\label{diamond_norm_thm}
Let $\map{E}$ be an $n$-qubit degree-$d$ quantum channel. There is an algorithm that learns $\map{E}$ in diamond distance with precision $\epsilon$ and failure probability $\delta$, using only single-qubit state preparation and measurements with
\begin{equation}
\scalingdiamond
\end{equation}
queries to $\map{E}$.
\end{theorem}
\begin{proof}
    The proof closely follows the one of Theorem~\ref{our_protocol_thm}. The only difference is that now we will need
    \begin{equation}
        M_{2} = \map{O}\!\left(\frac{C^d}{c^{4}\epsilon^{2}} \log\frac{1}{c^{2}\delta}\right).
    \end{equation}
    Then, using Lemma~\ref{diamond_bound} we can estimate the diamond distance as
    \begin{align}
        \norm{\map{E} - \map{E}''}_{\diamond}
    &\leq \sum_{\balpha,\bbeta\in X_{c}} |\chi_{\balpha,\bbeta} - \chi''_{\balpha,\bbeta}|
    + \sum_{\balpha\lor \bbeta\notin X_{c}} |\chi_{\balpha,\bbeta}| \\
    &\leq \epsilon 
       + \sqrt{2c}(4n)^d,
    \end{align}
    where we have used~\eqref{elem} to bound the second term along side the fact that a $d$-degree channel has at most $(4n)^d$ nonzero elements. Therefore by choosing $c=\frac{\epsilon^2}{2}(4n)^{-2d}$ we conclude the proof.    
\end{proof}

\subsection{Unitary acting on polylogarithmic number of qubits} \label{fixed_U}
Let us consider a unitary map $\map{U}$ that only acts on a set $Q$ of qubits of cardinality $q$. Then, given the product structure of the $\bG$ function, we will have that the rotated $\bG^{\map{U}}$ will be different only on the qubits on which $\map{U}$ is acting. For this we can write
\begin{equation}
    G^{\map{U}}_{\bgamma\bdelta}(\vr,\vs) = G_{\bgamma\bdelta}^{Q}\left(\prod_{i \in \bar{Q}} \ggd \right),
\end{equation}
where we have called $G_{\bgamma\bdelta}^{Q}$ the part of the function with support on $Q$ and $\bar{Q}$ the set of idle qubits. Similarly we know that $\chi_{\balpha\bbeta}^{\map{C}}$ will differ from $\chi_{\balpha\bbeta}^{\map{E}}$ on at most $q$ qubits for every $\balpha,\bbeta$ so it will have nonzero entries for Pauli strings that have weight at most $d + q$. So, writing down expression~\eqref{pseudo_variance} for this case we have
\begin{equation}
    \sum_{\balpha\bbeta} \chi_{\balpha\bbeta}^{\map{C}}\left( \sum_{\vr_Q,\vs_Q}  \left|G^{Q}_{\bgamma\bdelta}(\vr_Q,\vs_Q)\right|^2 \prod_{i\in Q} \fab \right) \left(\prod_{i\in \bar{Q}} \sum_{r_i,s_i }|\ggd|^2 \fab\right),
\end{equation}
where we have defined $\vr_Q,\vs_Q$ as the part of the input and output state that has support on $Q$. If we call $K$ the value of the first parenthesis, we can write from Eq.~\eqref{var_U}
\begin{align}
    \text{Var}(G^{\map{U}}_{\bgamma\bdelta}(\vr,\vs)) &\leq K \sum_{\balpha\bbeta} \chi_{\balpha\bbeta}^{\map{C}} \left(\prod_{i\in \bar{Q}} \sum_{\substack{r_i,s_i \\ i \in \bar{Q}}}|\ggd|^2 \fab\right) \\
    & \leq K C^{4d} \sum_{\balpha\bbeta} \chi_{\balpha\bbeta}^{\map{C}} \delta_{\balpha_{\bar{Q}}\bbeta_{\bar{Q}}} \\
    & \leq K C^{4d} \left(\sum_{\balpha_Q\bbeta_Q} \chi_{\balpha_Q\bbeta_Q}^{\map{C}}+1\right) \\
    & \leq (4^{q}+1)KC^{4d} \leq (K')^qC^{4d},
\end{align}
where from line one to two we have used once again that the number of qubits on which $\alpha_i,\beta_i,\gamma_i$ or $\delta_i$ differ from the identity is at most $4d$ on $\bar{Q}$ and we have used condition (ii) defined in Appendix~\ref{sample_complexity}. From line two to three we have used the fact that the trace of $\bchi$ is smaller equal than $1$ and from line three to four we have used the fact that the process matrix is positive semidefinite, so each off-diagonal term is bounded by the average of the corresponding diagonal terms ($\chi_{\balpha\bbeta} \leq (\chi_{\balpha\balpha}+\chi_{\bbeta\bbeta})/2$) so
\begin{equation}
    \sum_{\balpha\bbeta} \chi_{\balpha\bbeta} \leq \sum_{\balpha\bbeta} \chi_{\balpha\balpha} \leq 4^{q}.
\end{equation}
At last we defined $K'$ as $5^qK^{1/q}$ and got the final inequality.
We can now see that if we take $q = \map{O}(\text{poly}(\log(n)))$ we would get that the scaling of the variance becomes now polynomial in $n$, instead of constant
\begin{equation}
    \text{Var}(G^{\map{U}}_{\bgamma\bdelta}(\vr,\vs)) = \map{O}(e^qe^{d}) \rightarrow \map{O}(n^pe^{d}), 
\end{equation}
where $p$ is the product of the exponents appearing in the polynomial dependence on $\log$ of the support of such gates.
If we accept this trade-off we would be capable of performing efficient process-tomography on any gate acting on a polylogarithmic number of qubits, via Eq.~\eqref{sample_complexity_variance}.

\subsection{One layer of two-qubit gates}\label{layer}
As mentioned above we are interested in benchmarking the two-qubit gates of our device, so $\map{U}$ will be a layer of two-qubit gates. We will therefore define $n/2$ unitaries $\map{U}_{\bj}$ each one acting on a pair of qubits $\bj = (j_0,j_1)$. We will indicate with $\br_{\bj}$ and $\bs_{\bj}$ the qubits on which $\map{U}_{\bj}$ is acting.  We will call $J$ the set of pair of qubits such that no qubit is ever found in more than one pair at a time. This set has of course cardinality $n/2$.\\

As mentioned in the main text $G^{\map{U}}_{\balpha\bbeta}(\vr,\vs)$ can now be written as a product of $n/2$ two-qubit functions,
\begin{align}
    G^{\map{U}}_{\balpha\bbeta}(\vec{r},\vs) 
    & = \prod_{\bj\in J} g_{\map{U}_{\bj}}^{\balpha_{\bj}\bbeta_{\bj}} (\br_{\bj},\bs_{\bj}),
\end{align}
where we have defined the two qubit rotated $\vg$ function as $g_{\map{U}_{\bj}}^{\alpha_{\bj}\beta_{\bj}} (\br_{\bj},\bs_{\bj})$. Let us now dive into more details of the derivations of the minimization procedure. In this setting we can write expression~\eqref{pseudo_variance} as
\begin{equation} \label{var_2q}
     \frac{1}{18^n}\sum_{\balpha\bbeta} \chi_{\balpha\bbeta}^{\map{E}} \prod_{\bj\in J} \left(\sum_{\br_{\bj},\bs_{\bj}}|g_{\map{U}_{\bj}}^{\bgamma_{\bj}\bdelta_{\bj}} (\br_{\bj},\bs_{\bj})|^2 \bra{\br_{\bj}}\sigma_{\alpha_{\bj}}\map{U}_\bj(\ketbra{\bs_{\bj}})\sigma_{\beta_{\bj}}\ket{\br_{\bj}}\right).
\end{equation}
Since $\chi_{\balpha\bbeta}^{\map{E}}$ is nonzero only of for the Pauli operators with degree at most $d$, to have variance not scaling exponentially with system size, we would like the term in bracket to be bounded by $1$ for $\alpha_{\bj},\beta_{\bj},\gamma_{\bj},\delta_{\bj} = 0$,
\begin{align} \label{pseudo_var_U}
     \sum_{\br_{\bj},\bs_{\bj}}|g^{\map{U}_{\bj}}_{00} (\br_{\bj},\bs_{\bj})|^2 \bra{\br_{\bj}}\map{U}_{\bj}(\ketbra{\bs_{\bj}})\ket{\br_{\bj}} &= \sum_{\br_{\bj},\bs_{\bj}}|g^{\map{U}_{\bj}}_{00} (\br_{\bj},\bs_{\bj})|^2 p_{\map{U}_\bj}(\br_\bj,\bs_\bj) \\
     &= \sum_{\balpha_\bj\bbeta_{\bj}}\chi_{\balpha\bbeta}^{\map{U}_\bj}\sum_{\br_\bj,\bs_\bj}|g^{\map{U}_{\bj}}_{00} (\br_\bj,\bs_\bj)|^2 f_{\balpha_\bj\bbeta_\bj}(\br_\bj,\bs_\bj),
\end{align}
where we have defined $p_{\map{U}_\bj}(\br_\bj,\bs_\bj)$ as $(1/18^2)\bra{\br_{\bj}}\map{U}_{\bj}(\ketbra{\bs_{\bj}})\ket{\br_{\bj}} $. Now $f_{\balpha_\bj\bbeta_\bj}(\br_\bj,\bs_\bj) ,= f_{\alpha_{j_0}\beta_{j_0}}(r_{j_0},s_{j_0})\otimes f_{\alpha_{j_1}\beta_{j_1}}(r_{j_1},s_{j_1})$ acts on two qubits and can be thought as a $(1296\times 256)$ matrix, similarly to what done before. It is now possible to define the left kernel of $f_{\balpha_\bj\bbeta_\bj}$ as a $(1040 \times 1296)$ matrix $\mathbf{K}$ and again we will have that $g_{\map{U}_{\bj}}^{\bgamma_\bj\bdelta_\bj} + \vec{x}^{\mathrm{T}}K$ is the inverse of $f_{\balpha_\bj\bbeta_\bj}$. We are now left with minimizing~\eqref{pseudo_var_U} for the $1040$ free parameters encoded in $\vec{x}$,
\begin{align}
\min_{\vec{x}}\sum_{\br_\bj,\bs_\bj}|g_{\map{U}_{\bj}}^{00} + \vec{x}^{\mathrm{T}}K|^2 p_{\map{U}_\bj}(\br_\bj,\bs_\bj). 
\end{align}
This again is a quadratic minimization problem with a well known analytic solution. We find that this term is bounded by $1$ if $\map{U}$ is a Clifford gate.
\begin{equation}\label{var_bound_cl}
    \map{U} \in \text{Clifford} \quad \implies \quad \min_{\vec{x}}\sum_{\br_\bj,\bs_\bj}|g_{\map{U}_{\bj}}^{00} (\br_\bj,\bs_\bj) + \vec{x}^{\mathrm{T}}K|^2 p_{\map{U}_\bj}(\br_\bj,\bs_\bj) = 1.
\end{equation}
The code to get the minimized $g_{\map{U}_{\bj}}$ for a generic two qubit unitary is publicly available at Ref.~\cite{git}. In the same location it is possible to find the explicit solution for the \textit{iSWAP} gate as well as the \textit{CNOT}. 
Given this result let us now give an analytic proof of sample complexity. 

\subsection{Proof of sample complexity for a layer of two-qubit gates}\label{post_sample}
 When the bound in~\eqref{var_bound_cl} holds, we can bound the variance. Let us call $C$ the constant such that $\sum_{\br_{\bj},\bs_{\bj}}|g_{\map{U}_{\bj}}^{\bgamma_{\bj}\bdelta_{\bj}} (\br_{\bj},\bs_{\bj})|^2 \bra{\br_{\bj}}\sigma_{\alpha_{\bj}}\map{U}_\bj(\ketbra{\bs_{\bj}})\sigma_{\beta_{\bj}}\ket{\br_{\bj}} \leq C$ for every $\alpha_\bj,\beta_\bj,\gamma_\bj,\delta_\bj$. Then we rewrite Eq.~\eqref{var_U} as
\begin{align}
    \text{Var}(G_{\bgamma\bdelta}) &\leq \sum_{\balpha\bbeta} \chi_{\balpha\bbeta}^{\map{E}} \sum_{\vr,\vs} |G^{\map{U}}_{\bgamma\bdelta}(\vr,\vs)|^2 \bra{\vr}\sigma_{\balpha}\map{U} (\ketbra{\vs})\sigma_{\bbeta}\ket{\vr} \\
    &= \sum_{\balpha\bbeta} \chi_{\balpha\bbeta}^{\map{E}} \prod_{\bj\in J} \left(\sum_{\br_{\bj},\bs_{\bj}}|g_{\map{U}_{\bj}}^{\bgamma_{\bj}\bdelta_{\bj}} (\br_{\bj},\bs_{\bj})|^2 \bra{\br_{\bj}}\sigma_{\alpha_{\bj}}\map{U}_\bj(\ketbra{\bs_{\bj}})\sigma_{\beta_{\bj}}\ket{\br_{\bj}}\right) \\ \label{bound1}
    &\leq C^{4d} \sum_{\balpha\bbeta} \chi_{\balpha\bbeta}^{\map{E}},
\end{align}
where from line two to three we have used the fact that the indexes $\alpha_\bj,\beta_\bj,\gamma_\bj,\delta_\bj$ differ from the identity on at most $4d$ qubits and for the rest the term in bracket is equal to $1$. We are now left to bound the sum of the entries of the process matrix $\bchi$. To do that we will make use of Lemma~\ref{trace_noninc} and of the known properties of $p$-norms: $\norm{x}_r \leq n^{1/r-1/p}\norm{x}_p$ for $1\leq r < p$. By choorsing $r=1$ and $p=2d/(d+1)$ we get the following bound:
\begin{equation}\label{bound_pp}
    \sum_{\balpha\bbeta} \chi_{\balpha\bbeta}^{\map{E}} \leq  \norm{\bchi^{\map{E}}}_1 \leq n^{1-\frac{d+1}{2d}} \norm{\bchi^{\map{E}}}_{\frac{2d}{d+1}} \leq n K^d.
\end{equation}
Therefore we obtain
\begin{equation}\label{bound2}
    \text{Var}(G_{\bgamma\bdelta}) \leq nC^{4d} K^d.
\end{equation}
Making use of Eq.~\eqref{sample_complexity_variance} we are able to do efficient postprocessing of a layer of two-qubit Clifford gates with just a linear overhead. \par
If now we want to estimate the variance of the function $\bG_{\map{U}}$ corresponding to a layer of two-qubit Clifford gates and some non-Clifford gates, we need to put together the content of this section with Appendix~\ref{fixed_U}. Given the structure of the $\bG$ matrix we can split the function in its Clifford and non-Clifford component,
\begin{equation}
    \bG_{\map{U}} = \bG_{\map{U}_{Cl}} \cdot \bG_{\map{U}_{non-Cl}}.
\end{equation}
Then we can write
\begin{equation}
    \text{Var}(\bG_{\map{U}}) \leq \text{Var}(\bG_{\map{U}_{Cl}}) \cdot \text{Var}( \bG_{\map{U}_{non-Cl}}) \leq ne^qC^d,
\end{equation}
where $q$ is the number of qubits on which the non-Clifford gates are acting. Therefore as long as the number of qubits on which the non-Clifford gates are acting is polylogarithmic we have a variance of a single entry scaling polynomially in system size. \par
If we want to estimate the reconstructed channel in $\ell_2$-norm and diamond norm we need to adapt the proof of Theorem~\ref{our_protocol_thm} to take into account the different scaling of the variance. This brings as to state the following Lemma.
\begin{lemma}
    Let $\map{E}$ be an $n$-qubit degree-$d$ quantum channel, preceded by a gate layer composed of arbitrary unitaries with total support on at most $q$ qubits, and two-qubit Clifford gates acting on the remaining qubits. Let $\map{C}$ be the composition of the gate layer and $\map{E}$. There is an algorithm that learns $\map{E}$ in $\ell_{2}$-distance with precision $\epsilon$ and failure probability $\delta$, using only single-qubit state preparation and measurements with
    \begin{equation}
    \scalingeqppexp
    \end{equation}
    queries to $\map{C}$. Similarly there is an algorithm that learns $\map{E}$ in diamond distance with precision $\epsilon$ and failure probability $\delta$, using only single-qubit state preparation and measurements with
    \begin{equation}
    \scalingdiamondpp.
    \end{equation}
    
\end{lemma}
\begin{proof}
    The proof is a straightforward adaptation from Theorem~\ref{our_protocol_thm} and it only requires substituting $C^d$ with $ne^qC^d$. Now we would have $M_1 = \map{O}\!\left(\frac{ne^qC^d}{c^{2}} \log \frac{(4n)^d}{\delta}\right)$ and $M_2  = \map{O}\!\left(\frac{ne^qC^d}{c^{2}\epsilon^{2}} \log\frac{1}{c^{2}\delta}\right)$. This translates to a final sample complexity of
\begin{equation}
    M = \scalingeqppexp.
\end{equation}
Doing the same substitution in Theorem~\ref{diamond_norm_thm} we get the desired scaling,
    \begin{equation}
    M = \scalingdiamondpp.
    \end{equation}
Therefore we can see that restricting to a polylogarithmic number of non-Clifford gates, we would get $e^q \rightarrow n^p$ where $p$ is related to the degree of the polynomial in $\log$ number of non-Clifford gates. This preserves the efficiency of the protocol while introducing a polynomial factor in system size.
\end{proof}

\subsection{Numerical analysis}\label{post_num}
Since for the case of sampling directly of the error channel $\map{E}$ the scaling is constant for every entry of $\bchi$ we suspect that the linear bound that we get in this case is not tight. To verify that, we consider a layer of two qubit gates followed by the single-qubit noise model defined in~\eqref{err_ch_sq}. We then compute the variance of the function $\bG^{\map{U}}$, that in this case is a product of two qubit functions $\vg_{iSWAP}$.
In Fig.~\ref{fig:var} is reported the numerical computation of the variance of the function $G^{\map{U}}_{00}$ (that is the entry corresponding to all identities) for increasing system size. We can see that the value saturates to a constant, similar to what would happen without classical postprocessing. So, it is possible that even though the best bound we could find is linear, the scaling is actually constant in some relevant cases. \par
Given the previous results, we investigated under which conditions the chain of inequalities ceases to be tight. Specifically, we analyzed the behavior of the bounds going from Eqs.~\eqref{bound1} to~\eqref{bound2}. To this end, we generated random positive semidefinite matrices of increasing dimension. Specifically for every $n$ we constructed matrices of dimension $\sim (n^2\times n^2)$ to simulate the scaling with system size of a two-degree Pauli channel. We observed that the sum of their elements does not grow with the system size, as shown by the results summarized in Fig.\ref{fig:sum}. While we are aware that this behavior does not hold for all positive semidefinite matrices, our findings suggest that, in many practically relevant scenarios, the scaling may indeed remain constant. \par
Given the encouraging results of the previous section, where numerical computations exhibited better scaling with respect to analytical bounds, we  computed the variance of $\bG^{\min}$ for a layer of non-Clifford gates, while maintaining the same error model as in Sec.~\ref{numerics}. More specifically we used a layer of $T$ gates,
\begin{equation}
    T = \begin{pmatrix}
        1 & 0 \\
        0 & \exp\left(\frac{i\pi}{4}\right).
    \end{pmatrix}
\end{equation}
Results are reported in Fig.~\ref{fig:T} and show exponential scaling. 

\begin{figure}[h!]
    \centering
    \begin{minipage}{0.45\textwidth}
        \centering
        \includegraphics[width=\textwidth]{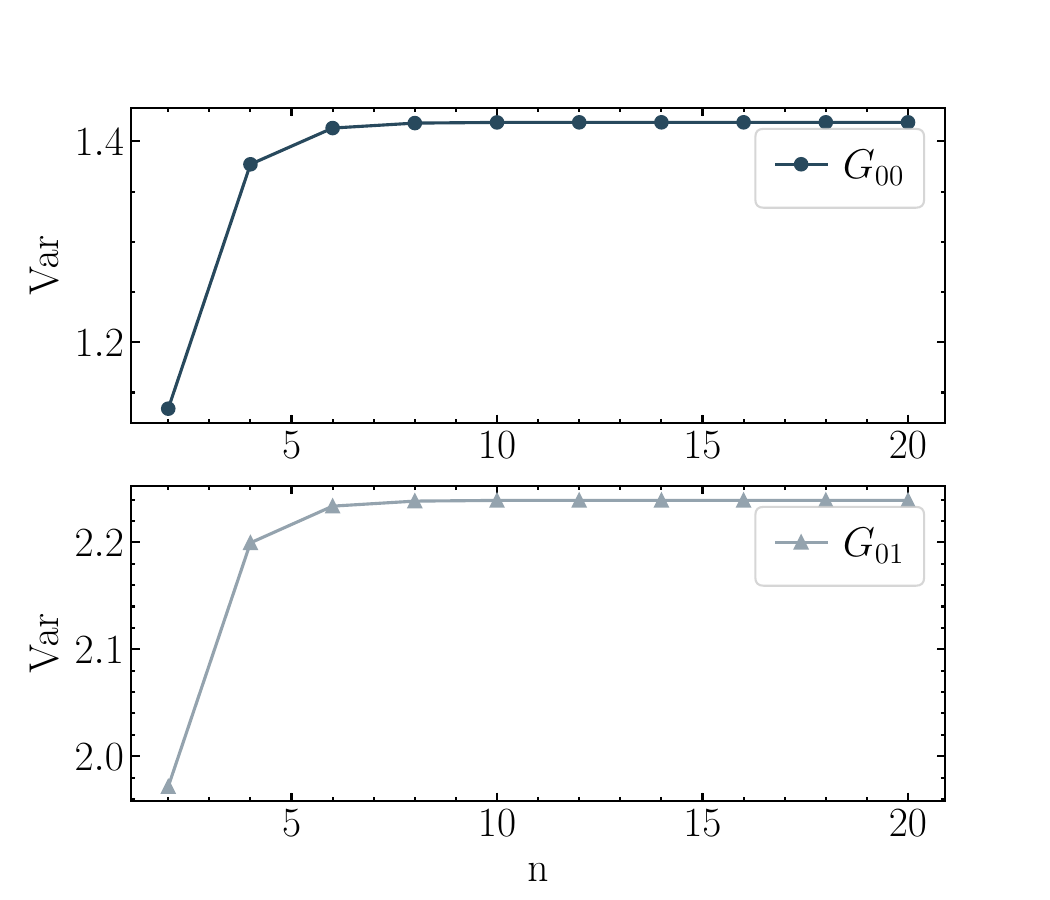}  
    \caption{Variance of the entry of $\bG$ when a layer of \textit{i}SWAP gates is applied, followed by an error channel with error parameters decaying exponentially as defined in Eq.~\eqref{err_ch_sq}. The computation is carried for the entries corresponding to $\balpha = \bbeta = (0,\ldots,0)$ (top panel) and $\balpha = (0,\ldots,0)$, $\bbeta = (x,0,\ldots,0)$ (bottom panel).}
    \label{fig:var} 
    \end{minipage}%
    \begin{minipage}{0.45\textwidth}
        \centering
        \includegraphics[width=\textwidth]{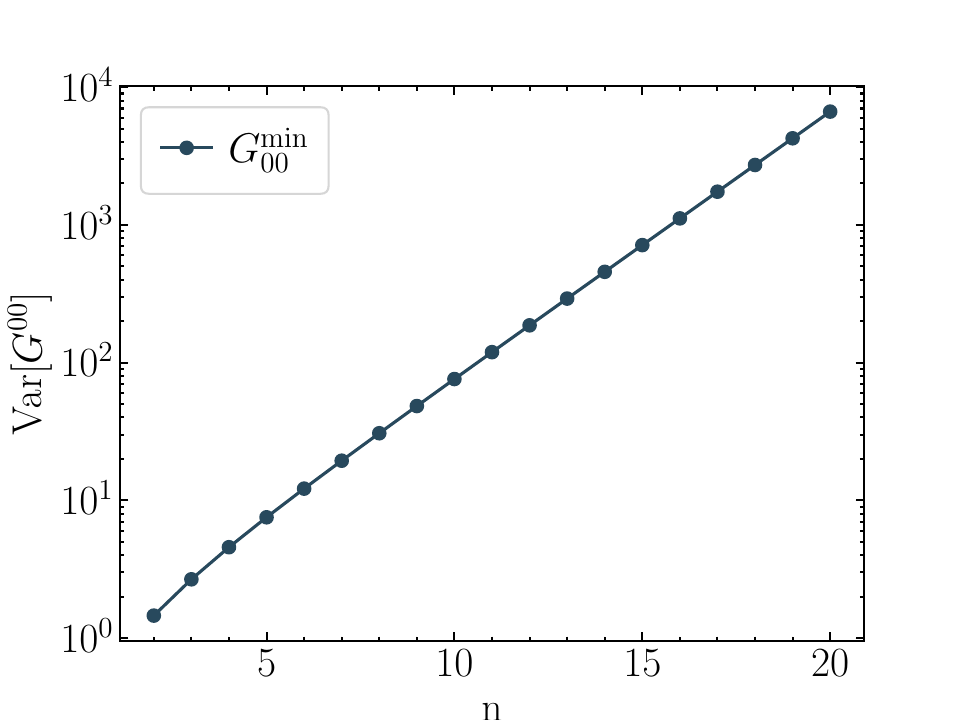}  
    \caption{Variance of the entry $G_{00}$ when a layer of \textit{T} gates is applied, followed by an error channel with error parameters decaying exponentially defined in~\eqref{err_ch_sq}.}
    \label{fig:T}  
    \end{minipage}
\end{figure}
\begin{figure}[h!]
    \centering
    \includegraphics[width=0.5\textwidth]{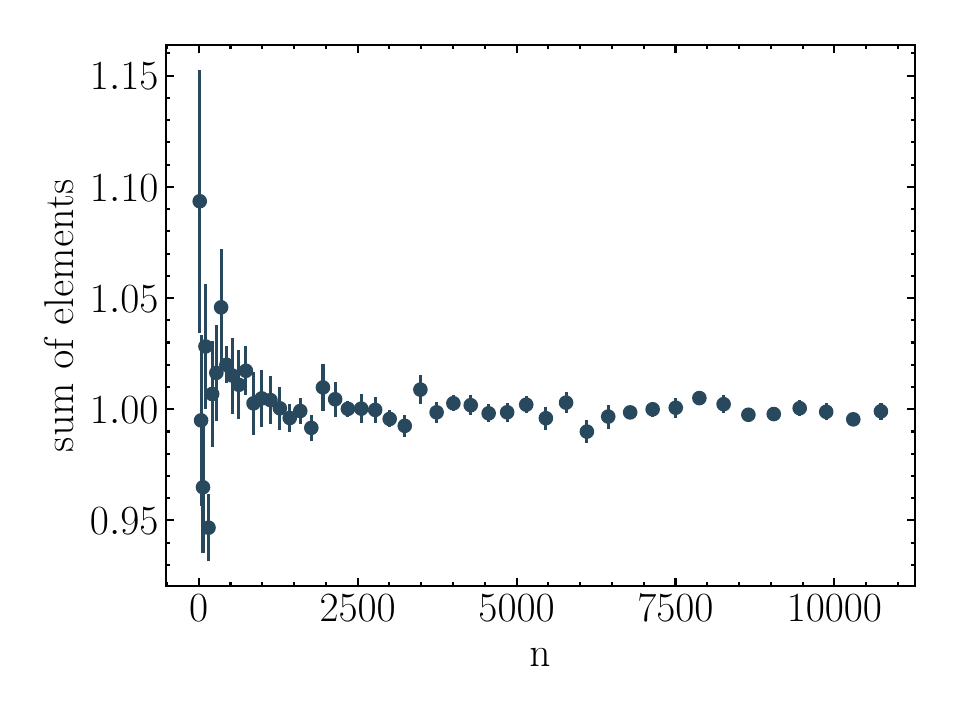} 
    \caption{Sum of elements of random positive semidefinite matrices for increasing system size. The average is taken over $10$ indipendent samples and the error bars are computed as the standard deviation.}
    \label{fig:sum}  
\end{figure}
\end{widetext}

\FloatBarrier
\newpage	
\bibliographystyle{unsrtnat}  
\bibliography{sample}

\end{document}